%% file: 0-qqQQmain.tex
\pdfoutput=1 
\documentclass[a4paper,11pt]{article}
\pdfoutput=1
\usepackage[english]{babel}
\usepackage{jheppub}
\usepackage{subfigure}
\usepackage{xspace}
\usepackage{amsmath,amssymb,textcomp,enumerate,alltt,xspace,multirow}
\usepackage{tensor}
\usepackage{slashed}
\usepackage{bbm}
\usepackage{multicol}

\usepackage{graphicx}
\usepackage{relsize}
\usepackage{pstricks}
\usepackage{color}
\usepackage{xcolor}
\graphicspath{fig}

\makeatletter
\g@addto@macro\bfseries{\boldmath}
\makeatother

\input texmacros.tex

\preprint{
\begin{flushright}
MSUHEP-21-016 \\
OUTP-21-20P   \\
\end{flushright}
}

\title{Three-loop helicity amplitudes for four-quark scattering in massless QCD}

\author[a,c]{Fabrizio Caola,}
\author[b]{Amlan Chakraborty,}
\author[a,d]{Giulio Gambuti,}
\author[b]{Andreas von Manteuffel,}
\author[a]{Lorenzo Tancredi}
\emailAdd{fabrizio.caola@physics.ox.ac.uk}
\emailAdd{chakra69@msu.edu }
\emailAdd{giulio.gambuti@physics.ox.ac.uk}
\emailAdd{vmante@msu.edu}
\emailAdd{lorenzo.tancredi@physics.ox.ac.uk}

\affiliation[a]{
Rudolf Peierls Centre for Theoretical Physics, University of Oxford,\\
Clarendon Laboratory, Parks Road, Oxford OX1 3PU
}

\affiliation[b]{
Department  of  Physics  and  Astronomy,  Michigan  State  University,\\
East  Lansing,  Michigan  48824,  USA
}

\affiliation[c]{
Wadham College, University of Oxford, Parks Road, Oxford OX1 3PN, UK
}

\affiliation[d]{
New College, University of Oxford, Holywell Street, Oxford OX1 3BN, UK
}

\abstract{
We compute the three-loop corrections to the helicity amplitudes for $q\bar{q}\to Q\bar{Q}$ scattering in massless QCD.
In the Lorentz decomposition of the scattering amplitude we avoid evanescent  Lorentz structures and map the corresponding form factors directly to the physical helicity amplitudes.
We reduce the amplitudes to master integrals and express them in terms of harmonic polylogarithms.
The renormalised amplitudes exhibit infrared divergences of dipole and quadrupole type, as predicted by previous work on the infrared structure of multileg scattering amplitudes.
We derive the finite remainders and present explicit results for all relevant partonic channels, both for equal and different quark flavours.
}

\keywords{QCD, Collider Physics, N3LO calculations}

\allowdisplaybreaks[1]

\begin{document}
\maketitle

\input{1-intro}
\input{2-notation}

\input{3-amplitude}
\input{4-computation}
\input{5-subtraction}
\input{6-results}

\input{7-other_processes}
\input{8-conclusions}

\section*{Acknowledgments}
LT is grateful to T.\ Peraro for
collaboration on a related project on 
the tensor decomposition in the 't Hooft-Veltman scheme used
in this paper. We thank F. Buccioni for his assistance in comparing our one-loop results to \texttt{OpenLoops}.
The research of FC is
partially supported by the ERC Starting Grant 804394
HipQCD. AvM is supported in part by the National Science Foundation through Grant 2013859.
LT and GG are supported by the Royal Society through Grant
URF/R1/191125.
\appendix

\input{9-appA}
\input{10-appB}
\input{11-appC}

\bibliography{biblio}

\end{document}

%% file: texmacros.tex
\newcommand{\ord}{\mathcal{O}}
\newcommand{\calO}{\ord}

\DeclareMathAlphabet\mathbfcal{OMS}{cmsy}{b}{n}

\DeclareMathOperator{\tr}{Tr}
\DeclareMathOperator{\sgn}{sgn}

\newcommand{\as}{\alpha_\mathrm{s}}
\newcommand{\asb}{\alpha_\mathrm{s,b}}



%% file: 1-intro.tex

\section{Introduction}

The success of the collider particle physics program, whose main
player today is the Large Hadron Collider (LHC) at CERN, relies
heavily on our ability to model with high precision and accuracy the
scattering of high energetic protons in Quantum Chromodynamics (QCD).
Thanks to asymptotic freedom and the factorization properties of QCD,
this intrinsically non-perturbative problem can be treated with
perturbative methods, supplemented by non-perturbative information about
the distribution of partons in the proton. Within this picture, an important
role is played by higher order perturbative QCD calculations, which allow
for a reliable and precise description of a wide range of collider processes
and observables.

Thanks to a concerted effort in the high-energy community over the
last few years, it is currently possible to compute predictions for
many interesting reactions to second order in the strong coupling
expansion, $i.e.$ to what is usually referred to as
next-to-next-to-leading order (NNLO). 
This has required, on the one hand, 
major advances in computational techniques
for multi-loop scattering amplitudes~\cite{Tkachov:1981wb,Chetyrkin:1981qh,Hodges:2009hk,Gluza:2010ws,Ita:2015tya,Larsen:2015ped,Bohm:2017qme,Badger:2016uuq,vonManteuffel:2014ixa,Peraro:2016wsq,Peraro:2019svx,Guan:2019bcx,Pak:2011xt,Abreu:2019odu,Heller:2021qkz,Kotikov:1990kg,Bern:1993kr,Remiddi:1997ny,Gehrmann:1999as,Papadopoulos:2014lla,Dixon:1996wi,Henn:2013pwa,Primo:2016ebd,Goncharov,Remiddi:1999ew,Goncharov:2001iea,Goncharov:2010jf,Brown:2008um,Ablinger:2013cf,Panzer:2014caa,Duhr:2011zq,Duhr:2012fh,Duhr:2019tlz},
which, notably, have recently made it possible to compute
various $2 \to 3$ processes up to two loops in QCD~\cite{Badger:2017jhb,Abreu:2017hqn,Abreu:2018aqd,Abreu:2018zmy,Abreu:2018jgq,Abreu:2019rpt,Abreu:2020cwb,Chicherin:2018yne,Chicherin:2019xeg,Chawdhry:2020for,DeLaurentis:2020qle,Chawdhry:2018awn,Abreu:2020xvt,Agarwal:2021grm,Badger:2021nhg,Abreu:2021fuk,Agarwal:2021vdh,Chawdhry:2021mkw,Badger:2021imn,Gehrmann:2015bfy,Papadopoulos:2015jft,Gehrmann:2018yef,Chicherin:2018mue,Chicherin:2020oor}.
On the other hand, the use of these amplitudes to perform phenomenological
studies for the relevant processes at NNLO~\cite{Chawdhry:2019bji,Kallweit:2020gcp,Chawdhry:2021hkp,Czakon:2021mjy} has required the
development of so-called subtraction or slicing frameworks~\cite{GehrmannDeRidder:2005cm,Czakon:2010td,Caola:2017dug,Magnea:2018hab,Herzog:2018ily,DelDuca:2016ily,Cacciari:2015jma,Catani:2007vq,Gaunt:2015pea,Boughezal:2015dva}
to properly deal with the intricate IR divergences that appear in QCD reactions. 

Beyond NNLO, predictions at third order in the
perturbative couplings, i.e.\ at N$^3$LO, are
known only for a handful of important LHC processes~\cite{Anastasiou:2015vya,Duhr:2019kwi,
Dulat:2018bfe,Mistlberger:2018etf,Dreyer:2016oyx,Dreyer:2018qbw,
Billis:2021ecs,Chen:2021isd,Chen:2021vtu}. 
In particular, N$^3$LO results are currently available only for reactions that
require at most three-point three-loop integrals.
Given the remarkable success of the
experimental program at the LHC, it is desirable to extend these
calculations to more complex processes. A particularly interesting
one is di-jet production.  In fact, jets are ubiquitous at
hadron colliders, so understanding their dynamics is of great
interest.
Moreover, di-jet production is the first massless $2\to2$
process that has a non-trivial colour structure. This makes it an
ideal ground for studying the structure of perturbative QCD. For
example, it is by now well known that when four or more coloured
partons interact, starting at the three-loop order, non-trivial
colour correlations can affect the pattern of IR divergences,
generating new structures~\cite{Almelid:2015jia} beyond the standard dipole
formula~\cite{Sterman:2002qn,Aybat:2006wq,Aybat:2006mz,Becher:2009cu,Gardi:2009qi,Becher:2009qa,Dixon:2009gx}. Also, the
non-trivial colour structure may create subtle violations of the
factorization framework that is at the very core of theoretical
predictions at hadronic
colliders~\cite{Catani:2011st,Forshaw:2012bi,Forshaw:2006fk,Becher:2021zkk}.
This makes jet production at hadronic colliders an extremely interesting
process to investigate at higher orders. 

A key ingredient for the study of jet production at N$^3$LO is
provided by the virtual three-loop corrections to the scattering
amplitudes for the production of two jets in massless QCD.  Modulo
crossings, there are three main partonic channels that need to be
computed: four-gluon scattering, the scattering of two quarks and two
gluons, and the scattering of four quarks.  All ingredients necessary
for the calculation of the two-loop QCD corrections to these processes have
been known for a long time~\cite{Smirnov:1999gc,Tausk:1999vh,Glover:2001af,Anastasiou:2002zn,Glover:2003cm}, which 
have made it possible to compute the relevant scattering amplitudes~\cite{Anastasiou:2000kg,Bern:2003ck,Glover:2004si,DeFreitas:2004kmi}. Also, in view of extending these
calculations to three loops, results for the two loop helicity
amplitudes up to order $\epsilon^2$ have been
obtained~\cite{Ahmed:2019qtg}.  For what concerns the three loop
results, instead, the relevant master integrals have been computed in
ref.~\cite{Henn:2020lye}, and have then been used to obtain the first
three loop results for $2 \to 2$ scattering amplitudes in
supersymmetric theories~\cite{Henn:2016jdu,Henn:2019rgj}.  More
recently also the first three loop corrections to the production of
two photons in full QCD have been obtained~\cite{Caola:2020dfu}.

In this paper, we move one step further and consider one of the three classes of partonic processes  listed above, 
namely the scattering of four massless quarks. This particular process is interesting not only
because it allows us, for the first time, to check the full structure of IR divergences at three loops in QCD, 
but also because it involves two external spinor structures. In fact, this property makes the use of the standard form factors
method for the calculation of the helicity amplitudes particularly cumbersome, due to the fact that the  $\gamma$-algebra does not
close in  $d$ space-time dimensions. For the calculation of the helicity amplitudes we then make use of
a different approach, recently described in refs~\cite{Peraro:2019cjj,Peraro:2020sfm}, which allows us to calculate the helicity amplitudes
in a simpler way, corresponding to the 't Hooft-Veltman scheme (tHV)~\cite{tHooft:1972tcz} for the processes considered here.\footnote{See also ref.~\cite{Heller:2020owb} for an application of similar ideas in the case of a chiral theory, and ref.~\cite{Chen:2019wyb} for an alternative approach.} 
In doing this, we also expose some
subtleties in the usual approach to compute helicity amplitudes in 
tHV with the standard form factor method.
The rest of the paper is organised as follows. 
We start in section~\ref{Kinematics} by establishing the notation for the calculation of the
fundamental partonic channel $q\bar{q} \to Q \bar{Q}$, from which all other channels can by obtained by crossing.
We continue in section~\ref{The Scattering Amplitude},
where we describe the colour and tensor decomposition  of the scattering amplitude, and show how to 
efficiently compute the helicity amplitudes without considering evanescent Lorentz structures.
In section~\ref{computation} we provide details on our computational set up,
and in section~\ref{subtraction} we discuss the renormalisation and the infrared structure of the three-loop scattering amplitudes.
In section~\ref{results} we discuss our final results for the main partonic channel.
In section~\ref{extra_results} we then explain how to obtain all other partonic channels from our calculation, both for quarks of equal and different flavour.
Finally we conclude in section~\ref{conclusions}.  In 
appendices~\ref{sec:appB} and~\ref{sec:appA}
we provide some details on the structure of infrared divergences up at three loops, in particular focusing on the explicit derivation of the quadrupole terms, which appear for the first time with the scattering of at least four coloured partons at three loops.
In appendix~\ref{sec:appC}, we review the analytical continuation of the amplitudes to different regions of phase space.

%% file: 2-notation.tex

\section{Kinematics and Notation}\label{Kinematics}
We consider the massless quark-antiquark scattering process 
\begin{equation}
  { q}(p_1) \;+ \;\bar { q}(p_2)  \; \longrightarrow \;  { \bar Q}(-p_3) \;+ \; {   Q}(-p_4) \; ,    \label{s_channel} 
\end{equation}
where $q$ and $Q$ are in general differently-flavoured quarks
and $p_1^2=p_2^2=p_3^2=p_4^2=0$. From
the master process in eq.~\eqref{s_channel} one can obtain all $2 \to
2$ quark scattering amplitudes with arbitrary initial states, including equal-flavour scattering
$q=Q$.  We discuss in detail how this can be achieved in section~\ref{extra_results}.\\ The minus signs appearing in the final-state
momenta imply that all momenta are taken to be incoming,
\begin{equation}
p_1^\mu + p_2^\mu  + p_3^\mu + p_4^\mu = 0.
\end{equation}
This choice is convenient when performing the required crossings to obtain
all the relevant partonic channels.
The kinematics of the process eq.~\eqref{s_channel} can be parametrized in
terms of Mandelstam invariants, defined as
\begin{align}
s
 = (p_1 + p_2)^2, \quad
t 
 = (p_1 + p_3)^2, \quad
u 
 = (p_2 + p_3)^2,
\quad u = -t-s \,.
\end{align}
We also find it convenient to define the dimensionless ratios 
\begin{equation}\label{variables}
x=-\frac{t}{s}\,, 
\quad\quad y = x|_{p_2 \leftrightarrow p_3} = -\frac{s}{t} , \quad\quad z = x|_{p_2 \leftrightarrow p_4} = -\frac{t}{u}.
\end{equation}
The variables $y$, $z$ are not needed for the computation of process
\eqref{s_channel}, but will be convenient to describe all the other
processes which can be derived from it using crossing symmetry.  In
terms of these variables the physical scattering region is given by $s >
0\,, \; t,u < 0$ which imply $0 < x < 1\,.$
For our calculation, we work in QCD with $n_f$ massless quarks and 
$N_c$ colours. We denote the
generators of the fundamental representation of $SU(N_c)$ 
by $(T^a)_{ij}$, with $\tr[T^aT^b] =
\frac12 \delta_{ab}$. 
Also, we indicate with $C_F$ and $C_A$ the quadratic Casimir constants.
They are defined through
$({T^a})_{ij} ({T^a})_{jk} =
C_F \delta_{ik}$
and 
$f_{acd}f_{bcd} = C_A\delta_{ab}$
where $f_{abc}$ are the 
structure constants.
These definitions imply
\begin{equation}
  C_F=\frac{N_c^2-1}{2N_c},~~~~ C_A = N_c,
\end{equation}
for the algebra of $SU(N_c)$.

%% file: 3-amplitude.tex

\section{The Scattering Amplitude}\label{The Scattering Amplitude}

\subsection{Lorentz Decomposition} 

We compute the scattering amplitude for the process \eqref{s_channel}
up to three loops in QCD, i.e.\ up
to order $\calO(\asb^4)$ , where $\asb$ is the bare strong coupling constant.  
We find it convenient to write the amplitude as 
\begin{equation}\label{definition_A_bar}
\tensor{\mathcal{A}}{_{i_1} _{i_2} _{i_3} _{i_4}} (q_1\bar q_2 \rightarrow \bar Q_3  Q_4)  =  ( 4 \pi \asb ) \;  \tensor{\bar{\mathcal{A}}}{_{i_1} _{i_2} _{i_3} _{i_4}}  \; ,
\end{equation}
where we have made the external quarks colour indices $i_j$ explicit.
The colour space for our process can be spanned
using the two tensor structures\footnote{We point out that $\mathcal C_1$ and
  $\mathcal C_2$ are not orthogonal.}
\begin{equation}\label{colour_structures}
\mathcal{C}_1 = {\delta}_{ i_1 i_4} {\delta}_{ i_2 i_3}
\;, \quad\quad  \mathcal{C}_2 = {\delta}_{ i_1 i_2} {\delta}_{ i_3 i_4}\;.
\end{equation}
By decomposing $\bar{\mathcal{A}}_{i_1i_2i_3i_4} $ with respect to the
$\mathcal{C}_j$, $j=1,2$ basis, we can write it as a vector in
colour space:
\begin{equation}
\tensor{\bar{\mathcal{A}}}{_{i_1} _{i_2} _{i_3} _{i_4}}  =  \bar{\mathcal{A}}^{[1]} \: \mathcal{C}_1  +  \bar{\mathcal{A}}^{[2]} \: \mathcal{C}_2        \quad \longrightarrow \quad \bar{\mathbfcal{A}}  = \begin{pmatrix}
\bar{\mathcal{A}}^{[1]} \\
\bar{\mathcal{A}}^{[2]}
\end{pmatrix} \; ,
\end{equation}
where we indicate all colour-space vectors in boldface.
It is important to notice that the components of the colour vector $\bar{\mathbfcal{A}}$ are independently gauge invariant,  since a gauge transformation cannot mix different colour contributions to the amplitude.  From now on we will work in the vector notation defined above, unless explicitly stated.\\

Turning to the spin structure of the process,  
we further decompose the scattering amplitude in terms of 
in terms of a basis of Lorentz structures (``tensors'') $T_i$
\begin{equation}\label{decomp_tensors}
\bar{\mathbfcal{A}}  = \sum_{i=1}^{N_L}  \mathbfcal{F}_i \;  T_i \;,
\end{equation}
where the $ \mathbfcal{F}_i$ are scalar \textit{form factors} that
only depends on the Mandelstam invariants and
$N_L$ is the number of elements of the basis of Lorentz structures.
In our calculation, we employ dimensional regularization to deal with
ultraviolet (UV) and infrared (IR) divergences. This makes the
decomposition eq.~\eqref{decomp_tensors} subtle.  Indeed,
if one
works in Conventional Dimensional Regularisation (CDR), one finds 
that, since the $\gamma$-algebra in $d$
dimensions does not close,
the number $N_L$ of independent structures depends on the loop
order~\cite{Glover:2004si}.  However, since we are ultimately
interested in computing helicity amplitudes in four dimensions, we
find it convenient to work in a scheme, where we can 
ignore evanescent Lorentz structures right from the start.
In this approach, it is possible to show
that the number of Lorentz structures which are physically
relevant is the same at any number of loops ($N_L = N$), and it
equals the number of independent helicity
amplitudes~\cite{Peraro:2019cjj,Peraro:2020sfm}.
Specifically, we consider all internal momenta and polarizations in $d$ dimensions, but restrict momenta and polarizations of the external quarks
to a $4$-dimensional subspace.
A convenient choice for the two independent Lorentz structures describing our process is~\cite{Peraro:2020sfm}:
\begin{align} \label{tensors}
T_1 &= {\bar u} (p_2)\: \gamma_\alpha \: { u} (p_1) \times {\bar u} (p_4) \: \gamma^\alpha \:{ u}(p_3) \; ,
\quad
T_2 = {\bar u} (p_2)\: \slashed p_3 \: { u} (p_1) \times   {\bar u} (p_4) \: \slashed p_2 \:{ u}(p_3) \;.
\end{align} 
These two Lorentz structures are then sufficient at any loop order.

In order to isolate the form factors from the rest of the amplitude, we define tensor projectors $P_i $ satisfying
\begin{equation}
P_i \cdot T_j = \delta_{ij} \; ,
\end{equation}  
where the dot products indicates the sum over the polarisations of the external quarks, $P_i \cdot T_j = \sum_{\rm pol} P_i T_j$.  
It then follows from eq.~\eqref{decomp_tensors}  that
$
 P_i \cdot \bar{\mathbfcal{A}}  = \mathbfcal{F}_i .
$ 
 For the choice eq.~\eqref{tensors}, the explicit form of the projectors is
 \begin{equation}
   \begin{split}
     P_1 &= \frac{1}{(d-3)4s^2} \; T_1^\dagger +  \frac{t-u}{(d-3)s^2tu}\;  T_2^\dagger \; ,\\
     P_2 &= \frac{t-u}{(d-3)s^2tu} \; T_1^\dagger + \frac{(d-4) s^2+ 2t^2 + 2 u^2}{(d-3) 4 s^2 t^2 u^2} \; T_2^\dagger \; .
   \end{split}
   \label{eq:p12}
 \end{equation}
We recall here that the main advantage of working with scalar form factors $\mathbfcal{F}_i $ 
is that, by construction, they only contain scalar integrals, 
because all the Lorentz tensor structure has been factorised out by the basis tensors ${T_i}$.

\subsection{Helicity Amplitudes} 
Ultimately, we are interested in computing the helicity amplitudes 
$\mathbfcal{A}_{\lambda_1 \lambda_2 \lambda_3 \lambda_4}$, 
for the process in eq.~  \eqref{s_channel},
where we indicate with $\lambda_j$ the helicity of the (anti)particle with momentum $p_j$.
Since the quarks are massless, helicity is conserved along the quark lines and there are only four different
possibilities that we need to consider
\begin{equation}
(\lambda_1,\lambda_2,\lambda_3,\lambda_4) = (+,-,+,-),(+,-,-,+),(-,+,+,-),(-,+,-,+) \; .  
\end{equation}
Moreover, the symmetries of the process allow us to compute only the first two helicity amplitudes $(+,-,+,-),(+,-,-,+)$ and then obtain the other two by acting on the result with a parity transformation, which flips the signs of the external helicities: 
\begin{equation}
(\lambda_1,\lambda_2,\lambda_3,\lambda_4)  \xrightarrow{P} (-\lambda_1,-\lambda_2,-\lambda_3,-\lambda_4) \; .
\end{equation}
In what follows, we will focus on the two independent configurations
$(\lambda_1,\lambda_2,\lambda_3,\lambda_4) = (+,-,+,-),(+,-,-,+)$.\\
We adopt the following definition for helicity states of spin-$\frac{1}{2}$ fermions
\begin{align}
  |p \rangle = {\ u} (p,+) = \left[ \frac{1}{2} (1 + \gamma_5) \right] \: {u}(p),  \quad \quad   |p] = {\ u} (p,-) = \left[\frac{1}{2} (1 - \gamma_5) \right] \: {u}(p),  \\[10pt]
   \langle p| = {\ \bar u} (p,+) = {\ \bar u}(p) \left[\frac{1}{2} (1 + \gamma_5) \right], \:  \quad \quad   [ p | = {\ \bar  u} (p,-) =  {\bar u}(p)\left[ \frac{1}{2} (1 - \gamma_5) \right],
\end{align}
where we use the well known \textit{spinor-helicity formalism}~\cite{Dixon:1996wi}, and indicate with $\pm$ the projection of the (anti-)particle spin along its four-momentum.

Using this notation and starting from the general structure of the amplitude given in eqs~\eqref{decomp_tensors} and~\eqref{tensors},
we obtain 
\begin{align}
\bar{\mathbfcal{A}}_{+-+-}^{\scriptstyle q \bar q\rightarrow \bar Q Q} &=  
\mathbfcal{H}_1 \frac{\langle 13 \rangle}{\langle 24 \rangle} \,, \quad 
\bar{\mathbfcal{A}}_{+--+}^{\scriptstyle q \bar q\rightarrow \bar Q Q} =  
\mathbfcal{H}_2 \frac{\langle 14 \rangle}{\langle 2 3 \rangle}, \label{H_ij}
\end{align}
where 
\begin{equation}\label{H1H2def}
\mathbfcal{H}_1 = 2t \mathbfcal{F}_1  - tu \mathbfcal{F}_2,\quad \mathbfcal{H}_2 = 2u \mathbfcal{F}_1  + tu \mathbfcal{F}_2.
\end{equation}
In eqs.~\eqref{H_ij} we have introduced an explicit label for the 
process that we are considering, which will turn out to be useful later on when
we describe the other partonic channels.

Since we expect that final analytic results for the 
helicity amplitudes should display the maximum degree of simplicity,
in what follows we focus directly on the two 
linear combinations $\mathbfcal{H}_1$ and $\mathbfcal{H}_2$.
We write their expansion in terms of the bare coupling $\asb$ as
\begin{equation}\label{full_amplitude}
\mathbfcal{H}_i =  \mathbfcal{H}^{(0)}_i +\left(  \frac{\asb }{4 \pi}  \right)\: \:\mathbfcal{H}^{(1)}_i+\left(  \frac{\asb }{4 \pi}  \right)^2\:  \: \mathbfcal{H}^{(2)}_i+\left( \frac{\asb }{4 \pi}  \right)^3 \:  \:\mathbfcal{H}^{(3)}_i + \calO\left( \asb^4 \right).
\end{equation} 
In the next section, we discuss the computation of  $\mathbfcal{H}_1$ and $\mathbfcal{H}_2$ up to three loops in QCD.

%% file: 4-computation.tex

\section{Computation}
\label{computation}

We perform our calculations in dimensional regularization with $d = 4 - 2 \epsilon$ dimensions for all internal momenta and gluon fields. UV and IR singularities will then manifest themselves as poles in the dimensional regulator $\epsilon$.
In order to compute the  helicity amplitudes $\mathbfcal{H}_1$ and $\mathbfcal{H}_2$,  we begin by producing all relevant Feynman diagrams for the process in eq.~\eqref{s_channel} with \texttt{QGRAF}~\cite{Nogueira:1991ex}. Only 1 diagram contributes at tree level,  9 diagrams at one loop, 158 diagrams at two loops and 3584 at three loops. We give a few representative samples of the three-loop diagrams in figure~\ref{diagrams}. 
\begingroup
\centering
\begin{figure}[htbp]
\centering
\subfigure[]{\includegraphics[width=1.3in]{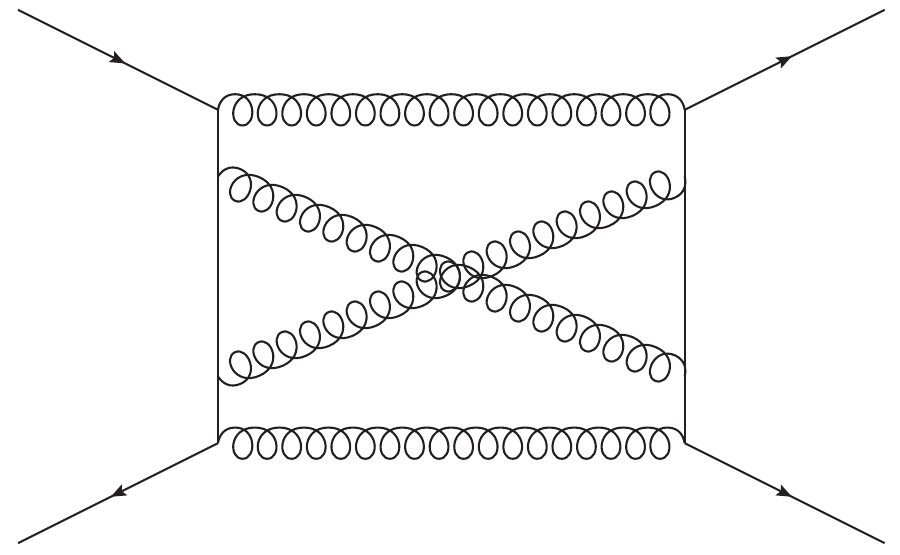}}\label{fig:1a}
\subfigure[] {\includegraphics[width=1.3in]{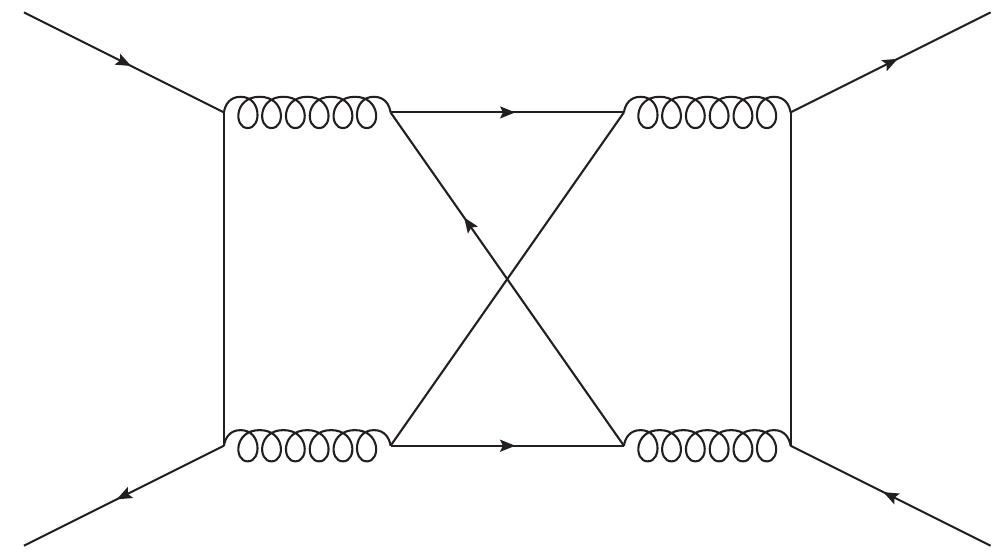}}\label{fig:1b}
\subfigure[]{\includegraphics[width=1.3in]{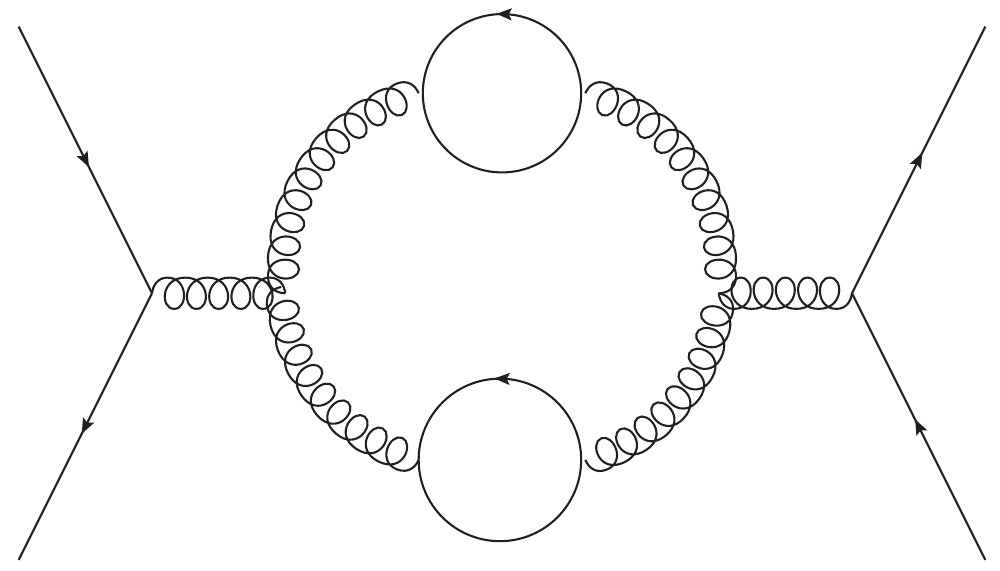}}\label{fig:1c}

\subfigure[]{\includegraphics[width=1.3in]{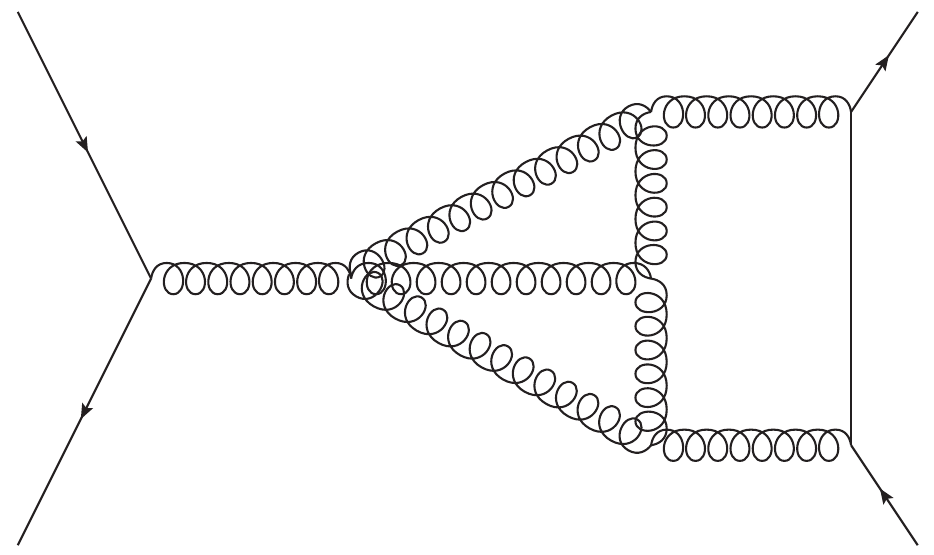}}\label{fig:1d}
\subfigure[] {\includegraphics[width=1.3in]{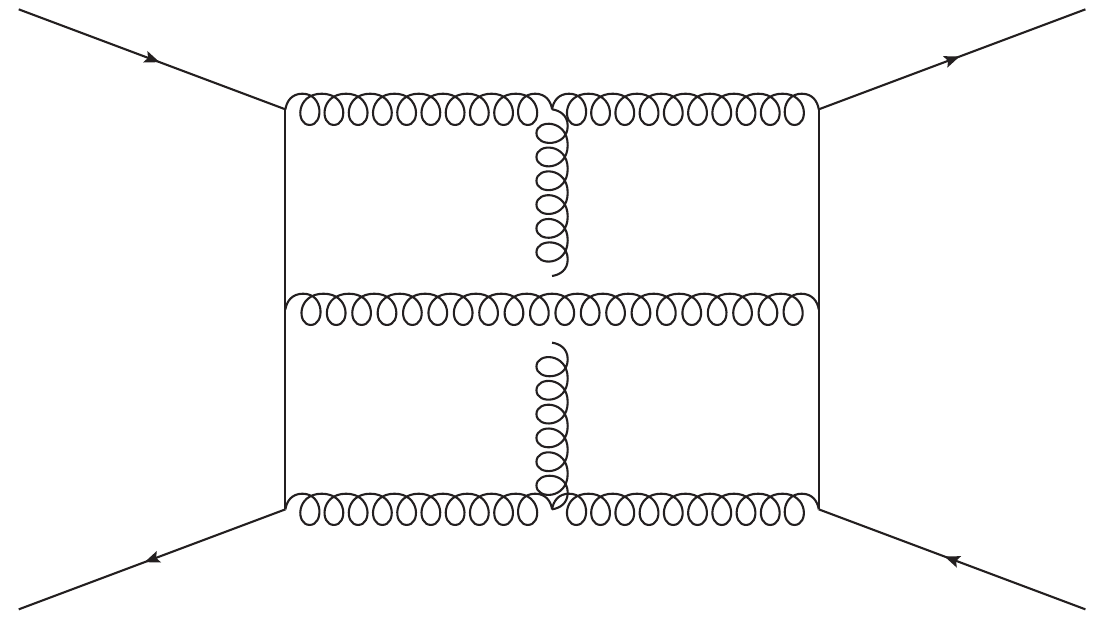}}\label{fig:1e}
\subfigure[]{\includegraphics[width=1.3in]{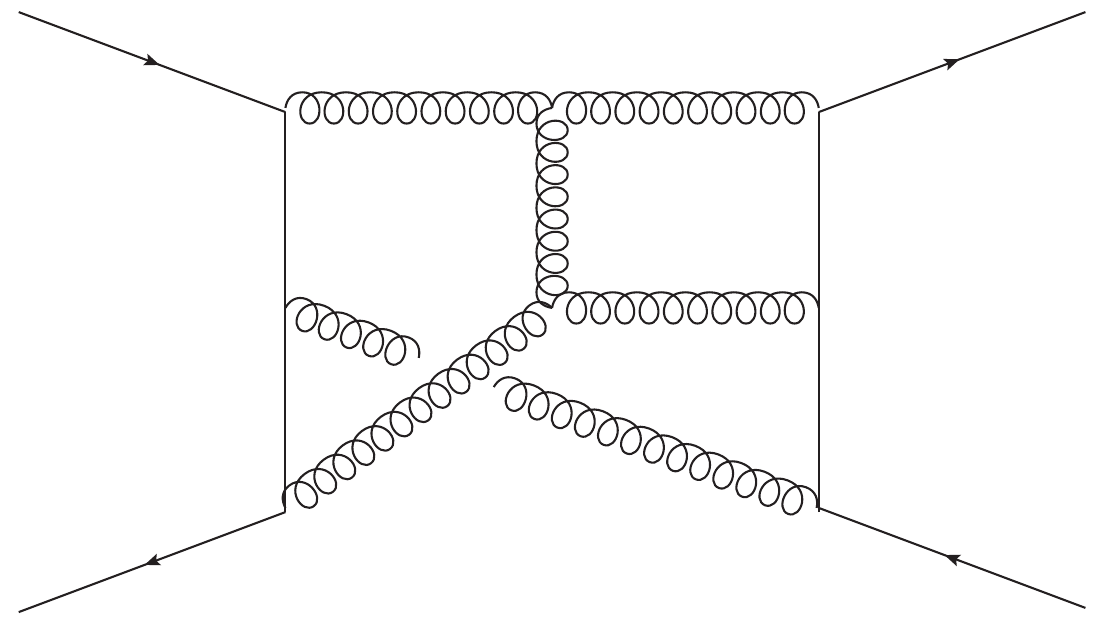}}\label{fig:1f}
\caption{Sample three loop diagrams
contributing to the process $q\bar q \rightarrow Q \bar Q $.} \label{diagrams}
\end{figure}
\endgroup
We use \texttt{FORM} \cite{Vermaseren:2000nd} to apply the tensor projectors of eq.~\eqref{eq:p12} to the diagrams, perform the Dirac traces and the colour algebra.  The latter can be boiled down to a repeated application of the identities

\begin{align}
  {({T^a})}_{ij} {({T^a})}_{kh} &= \frac12
  \left({\delta}_{ih} {\delta}_{kj} - \frac{1}{N_c} {\delta}_{ij} {\delta}_{kh}\right)\,, \quad
f^{abc} = - 2 \: i \: {\rm Tr}(T^a[T^b,T^c])\,
\end{align}
After the colour algebra  has been performed,  the quark colour indices can only appear via the two independent rank-4 tensor structures defined in \eqref{colour_structures}.   They appear in the amplitude accompanied by coefficients of the type 
\begin{equation}
n_f^a \: N_c^b \quad\text{with~}    a=0,\ldots,3,~ b=-4,\ldots,3\,.
\end{equation} 
Terms in the amplitude with different $a$ and $b$ are separately gauge invariant so there cannot be any gauge cancellations among them.
Because of this, we compute them separately, which allows
us to deal with smaller expressions.

After performing the colour and Dirac algebras we can express the helicity amplitudes as linear combinations
of scalar Feynman integrals with rational coefficients depending on the Mandelstam invariants $s$,  $t$ and the dimensional
regulator $\epsilon$. 
At $L$ loops, we write the integrals appearing in the amplitudes as
\begin{equation}\label{integrals}
\mathcal{I}^\text{top}_{n_1,n_2,...,n_N} = \mu_0^{2L\epsilon} e^{L \epsilon \gamma_E}  \int \prod_{i=1}^L \left( \frac{\mathrm{d}^d k_i}{i \pi^{\frac{d}{2}}} \right) \frac{1}{D_1^{n_1}D_2^{n_2} \dots D_N^{n_N}}
\end{equation}
where $\gamma_E \approx 0.5772$  is the Euler constant and $\mu_0$ is the dimensional regularisation scale. Here the factor $e^{L \epsilon \gamma_E}$ is purely conventional and it is chosen for later convenience, while the factor $\mu_0^{2L\epsilon}$ ensures that the integrals have integer mass dimension.
In general,  for a given process with $E$ independent external momenta and $L$ loops one needs $L(L+1)/2 + L E $ independent denominators to describe all possible scalar products of loop momenta with loop or external momenta.  
In our case $E=3$, and therefore we need $4$ denominators at one loop,  $9$ denominators at two loops and $15$ 
at three loops. A specific complete set of $\{D_i\}$ at a given loop order
is usually referred to as
an ``integral family''. 
Having in mind the calculation of the three-loop scattering amplitudes, it is useful to organise the relevant integrals 
in as few integral families as possible, up to permutation of the external momenta. 
While one family is sufficient at one loop, we need one planar family and one non-planar one at two loops 
and one
planar and two non-planar ones at three loops. 
We report them in Tabs\eqref{table:1}, \eqref{table:2} and \eqref{table:3} .
There, we indicate the loop momenta with $k_1$, $k_2$ and $k_3$.  We name PL the families corresponding to planar graphs and 
NPL,  NPL1,  NPL2 the ones corresponding to non-planar graphs. 
The different number of loop momenta for the planar topologies eliminates any ambiguity
in the naming convention.

\begin{table}
\centering
\begin{tabular}{ c || c  }
Family & PL \\
\hline\hline
$D_1 $& $k_1^2$  \\ 
$D_2 $& $(k_1 - p_1)^2$  \\  
$D_3 $& $(k_1 - p_1-p_2)^2$    \\
$D_4 $& $(k_1 - p_1-p_2-p_3)^2$  
\end{tabular}
\caption{Planar 1-loop integral family.}
\label{table:1}
\end{table}

\begin{table}
\centering
\begin{tabular}{ c || c | c }
Family & PL & NPL\\
\hline\hline
$D_1  $& $k_1^2$  &   $k_1^2$ \\ 
$D_2 $& $k_2^2$ &  $k_2^2$ \\  
$D_3 $& $(k_1 - k_2)^2$  &  $ (k_1-k_2)^2$  \\
$D_4 $& $(k_1 - p_1)^2$ &  $ (k_1-p_1)^2$ \\
$D_5 $& $(k_2 - p_1)^2$ &  $(k_2-p_1)^2$ \\
$D_6$& $(k_1 - p_1-p_2)^2$ &  $(k_1 - p_1-p_2)^2$ \\
$D_7 $& $(k_2 - p_1-p_2)^2$ &  $(k_1 - k_2+ p_3)^2$ \\
$D_8$ & $(k_1 - p_1-p_2-p_3)^2$ &  $(k_2 - p_1-p_2-p_3)^2$ \\
$D_9$ & $(k_2 - p_1-p_2-p_3)^2$ &  $(k_1-k_2-p_1-p_2)^2$ \\
\end{tabular}
\caption{Planar and non-planar 2-loop integral families.}
\label{table:2}
\end{table}

\begin{table}
\centering
\begin{tabular}{ c || c | c |c}
Family & PL & NPL1 & NPL2\\
\hline\hline
$D_1  $& $k_1^2$  & $k_1^2$ & $k_1^2$ \\ 
$D_2 $& $k_2^2$ & $k_1^2$ & $k_1^2$ \\  
$D_3 $& $k_3^2$ & $k_3^2$  & $k_3^2$    \\
$D_4 $& $(k_1 - p_1)^2$ & $(k_1 - p_1)^2$ & $(k_1 - p_1)^2$ \\
$D_5 $& $(k_2 - p_1)^2$ & $(k_2 - p_1)^2 $ & $(k_2 - p_1)^2$   \\
$D_6$& $(k_3 - p_1)^2$ & $(k_3 - p_1)^2$ & $(k_3 - p_1)^2$ \\
$D_7 $& $(k_1 - p_1-p_2)^2$ & $(k_1 - p_1-p_2)^2$ & $(k_1 - p_1-p_2)^2$ \\
$D_8$ & $(k_2 - p_1-p_2)^2$ & $(k_2 - p_1-p_2)^2$  & $(k_3 - p_1-p_2)^2$  \\
$D_9$ & $(k_3 - p_1-p_2)^2$ & $(k_3 - p_1-p_2)^2$ & $(k_1 - k_2)^2$ \\
$D_{10}$& $(k_1 - p_1-p_2-p_3)^2$ & $(k_1 - p_1-p_2-p_3)^2$ & $(k_2 - k_3)^2$ \\
$D_{11}$ & $(k_2 - p_1-p_2-p_3)^2$ & $(k_2 - p_1-p_2-p_3)^2$  & $(k_1 -k_2 -p_3)^2$  \\
$D_{12}$ & $(k_3 - p_1-p_2-p_3)^2$ & $(k_3 - p_1-p_2-p_3)^2$ & $(k_2 - k_3+p_1+p_2+p_3)^2$ \\
$D_{13}$ & $(k_1 - k_2)^2$ & $(k_1 - k_2)^2$ & $(k_2 + p_3)^2$\\
$D_{14}$ & $(k_1 - k_3)^2$ & $(k_2 - k_3)^2$  & $(k_1 - k_3)^2$  \\
$D_{15}$ & $(k_2 - k_3)^2$ & $(k_1 - k_2 + k_3)^2$ & $(k_2 - p_1 - p_2)^2$ \\
\end{tabular}
\caption{Planar and non-planar 3-loop integral families.}
\label{table:3}
\end{table}

As it is well known, the integrals in eq.~\eqref{integrals} are not
all independent and, instead, various types of relations can be
established among them, most notably by the algorithmic exploitation
of symmetry relations and integration-by-parts
identities~\cite{Tkachov:1981wb,Chetyrkin:1981qh}.  The latter in
particular allow one to reduce the number of independent integrals by
potentially several orders of magnitudes and to express the amplitude in terms of
a relatively small number of so-called master integrals.  While the
reduction to master integrals for $2 \to 2$ massless scattering up to
two loops can be performed very easily with automated tools, going one
order higher involves a considerable increase in complexity.  In
practice, at three loops we proceed as follows. Once the amplitude has
been expressed in terms of scalar integrals, we first use
\texttt{Reduze 2}~\cite{Studerus:2009ye,vonManteuffel:2012np} to find
trivially vanishing integrals, non-trivial symmetry relations among
the various integrals and corresponding ones obtained by permutations
of the external momenta.  This step already reduces the size of the
expressions significantly as well as the number of integrals within them.  We then use
\texttt{Finred}, an in-house implementation of the Laporta
algorithm~\cite{Laporta:2001dd} augmented by the use of finite-field
arithmetics~\cite{vonManteuffel:2014ixa,
  vonManteuffel:2016xki,Peraro:2016wsq,Peraro:2019svx} and
syzygy-based
techniques~\cite{Gluza:2010ws,Schabinger:2011dz,Ita:2015tya,Larsen:2015ped,Bohm:2017qme,Agarwal:2020dye},
in order to solve the system of integration-by-parts
identities satisfied by the remaining integrals.
This allows us to express the three-loop helicity
amplitudes for this process in terms of the basis of master integrals
computed in ref.~\cite{Henn:2020lye}. In this way we find that the
physical three-loop scattering amplitudes can be expressed in terms of
the same $486$ master integrals necessary for the calculation of
diphoton production at three loops~\cite{Caola:2020dfu}.

%% file: 5-subtraction.tex

\section{Ultraviolet and Infrared Subtraction}
\label{subtraction}

The result of the computation described in the previous section are the divergent helicity amplitudes for the process \eqref{s_channel} to $\mathcal{O}(\asb^4)$.  
In the following, we describe the UV renormalisation and IR subtraction of the divergent amplitude for this process. 

\subsection{Ultraviolet Renormalisation}
We work in massless QCD with an arbitrary number of fermion flavours $n_f$.  We adopt the standard $\overline{\mathrm{MS}}$ renormalisation scheme,  where the bare coupling $\asb$ is written in terms of the renormalised coupling $\as(\mu)$ in the following way
\begin{align}
\label{bare_to_phys}
\asb \: \mu_0^{2\epsilon} \: S_\epsilon &= \as \: \mu^{2\epsilon} Z[\alpha_s(\mu)] 
\end{align}
where $S_\epsilon = (4 \pi)^{-\epsilon} e^{-\gamma_E \epsilon}$, $\mu$ is the renormalisation scale (for the rest of the paper we set $\mu_0=\mu$) and
\begin{align}
Z[\alpha_s]  =  1 -  \left( \frac{\as}{4 \pi} \right) \frac{ \beta_0 }{\epsilon } + \left(\frac{\as}{4 \pi}\right)^2 \left( \frac{\beta_0^2}{\epsilon^2} - \frac{\beta_1 }{2 \epsilon} \right)- 
\left(\frac{\as}{4 \pi}\right)^3  \left( \frac{\beta_0^3}{\epsilon^3} - \frac{ 7}{6} \frac{\beta_0 \beta_1}{\epsilon^2}+ \frac{\beta_2}{3 \epsilon} \right)  + \mathcal{O}(\as^4)\,.
\end{align}
The $\beta$-function coefficients are defined through
\begin{align}
\frac{d \as }{d \log \mu} &= \beta(\as,\epsilon) = \beta(\as) - 2\epsilon \as \; , \qquad
\beta(\as) = -2 \as \sum_ {n=0}^\infty \beta_n \left(\frac{\as}{4 \pi}\right)^{n+1} \; ,
\end{align}
where in this equation $\as\equiv\as(\mu)$. To the relevant order, they read
\begin{align}
\beta_0 &= \frac{11}{3} C_A - \frac{2}{3}\: n_f \; , \nonumber\\
\beta_1 &= \frac{1}{3} \left(34 \: C_A^2-10\: C_A \:n_f \right)-2 \:C_F\: n_f \; ,\\ 
\beta_2 &= -\frac{1415 \:C_A^2 \:n_f}{54}+\frac{2857\: C_A^3}{54}-\frac{205\: C_A\: C_F\:
   n_f}{18}+\frac{79 \:C_A \:n_f^2}{54}+C_F^2 \:n_f+\frac{11\: C_F \:n_f^2}{9} \; .\nonumber
\end{align}
By inserting  eq.~\eqref{bare_to_phys} in the $\as$ expansion for the helicity amplitudes~\eqref{full_amplitude}, we obtain the renormalised helicity amplitudes
\begin{align}
\mathbfcal{H}_{i,\text{ren}}^{(0)} &=  \mathbfcal{H}_i^{(0)} \, , \\ 
\mathbfcal{H}_{i,\: \text{ren}}^{(1)} & =\mathbfcal{H}_i^{(1)}- \frac{\beta_0 }{\epsilon}   \mathbfcal{H}_i^{(0)} \, , \\ 
\mathbfcal{H}_{i,\: \text{ren}}^{(2)} &= \mathbfcal{H}_i^{(2)}  - \frac{2 \beta_0 }{\epsilon}  \mathbfcal{H}_i^{(1)}  + \frac{ \left(2 \beta_0^2- \beta_1\epsilon\right)}{2 \epsilon^2} \mathbfcal{H}_i^{(0)} \, ,\\ 
\mathbfcal{H}_{i,\: \text{ren}}^{(3)} & = \mathbfcal{H}_i^{(3)} -\frac{3 \beta_0}{\epsilon}    \mathbfcal{H}_i^{(2)} +\frac{ \left(3 \beta_0^2-\beta_1 \epsilon\right)}{\epsilon^2}  \mathbfcal{H}_i^{(1)} + \frac{ \left(7 \beta_1 \beta_0
   \epsilon -6 \beta_0^3-2 \beta_2 \epsilon^2 \right)}{6 \epsilon^3} \mathbfcal{H}_i^{(0)}  \; .
\label{hel_ampls_ren}
\end{align}
These are free from UV poles, but they still contain poles in $\epsilon$ of IR origin. We discuss how to subtract them in the next subsection.

\subsection{Infrared Subtraction}
While the structure of IR singularities of scattering
amplitudes in massless QCD up to two-loop order has been known for a
long time~\cite{Catani:1998bh}, its generalisation to three- and
higher-loop order has been understood only more recently, in
particular in the case where four or more coloured partons participate
to the scattering
process~\cite{Becher:2009cu,Becher:2009qa,Almelid:2015jia}.  In
particular, it has been shown that IR singularities are in one-to-one
correspondence to the UV poles of operator matrix elements in
SCET~\cite{Becher:2009cu,Becher:2009qa}.  Therefore, UV
renormalisation in SCET corresponds to IR subtraction in QCD and one
can write the finite remainder of the scattering amplitude by means of
a multiplicative colour-space operator $\mathbfcal{Z}$
as
\begin{equation}\label{Zeta_H}
\mathbfcal{H}_{i,\:\text{fin}} (\epsilon,\{p\})= \lim_{\epsilon \rightarrow 0} \mathbfcal{Z}^{-1}(\epsilon,\{p\},\mu) \; \mathbfcal{H}_{i,\:\text{ren}}(\epsilon,\{p\}) \; ,
\end{equation}
where $\{p\}$  stands for the dependence on the external kinematics. 
We point out that,  since we are working with vectors in colour-space as defined in Section \ref{The Scattering Amplitude}, the $\mathbfcal{Z}$ colour operator can be represented as a 2 by 2 matrix which mixes the colour structures defined in eq.~\eqref{colour_structures}. 
Solving a renormalisation group equation one finds that $\mathbfcal{Z}$ can be rewritten as 
\begin{equation}\label{exponentiation}
\mathbfcal{Z} (\epsilon,\{p\},\mu) = \mathbb{P} \exp \left[ \int_\mu^\infty \frac{\mathrm{d} \mu'}{\mu'}  \mathbf{\Gamma}(\{p\},\mu')\right] = \sum_{n=0}^{\infty} \left( \frac{\as}{4 \pi} \right)^n \mathbfcal{Z}_n \; ,
\end{equation}
with $\mathbb{P}$ the \textit{path-ordering} symbol, $i.e.$ operators are ordered from left to right with decreasing values of $\mu'$.
Following the notation of \cite{Almelid:2015jia}, where $\mathbf{\Gamma}$ was first computed up to three loops,  the anomalous dimension operator for 4 coloured external particles is written as 
\begin{equation}\label{anomalous_operator}
\mathbf{\Gamma}(\{p\},\mu) =  \mathbf{\Gamma}_{\text{dipole}}(\{p\},\mu)  + \mathbf{\Delta}_4 (\{p\}) \; .
\end{equation}
Above,  $ \mathbf{\Gamma}_\text{dipole}$ represents the well known dipole colour correlations between two coloured external legs, namely 
\begin{equation}\label{dipole}
\mathbf{\Gamma}_{\text{dipole}}(\{p\},\mu)  =  \sum_{1\leq i < j \leq 4} \mathbf{T}^a_i \; \mathbf{T}^a_j \; \gamma^\text{cusp}(\as) \; \log\left(\frac{\mu^2}{-s_{ij}-i \delta}\right)  \; + \; \sum_{i=1}^4 \; \gamma^i(\as) \; ,
\end{equation}
with $\mathbf{T}^a_i$ the $i$-th particle $SU(N_c)$ generator and from now on we use the shorthand $\alpha_s = \alpha_s(\mu)$ to indicate the renormalised coupling at scale $\mu$.
The constants $\gamma^\text{cusp}$~\cite{Korchemsky:1987wg,Moch:2004pa,Vogt:2004mw,Grozin:2014hna,Henn:2019swt,Huber:2019fxe,vonManteuffel:2020vjv}
and $\gamma^i$~\cite{Ravindran:2004mb,Moch:2005id,Moch:2005tm,Agarwal:2021zft}
are given in Appendix \ref{sec:appA}.  It is also useful to define the expansions 
\begin{equation}
\mathbf{\Gamma}_{\text{dipole}} = \sum_{n=0}^\infty \mathbf{\Gamma}_n \;  \left( \frac{\as}{4 \pi} \right)^{n+1}\,, \quad  \quad \Gamma '= \frac{\partial \mathbf{\Gamma}_{\text{dipole}}}{\partial \log \mu} = \sum_{n=0}^\infty \Gamma'_n \;  \left( \frac{\as}{4 \pi} \right)^{n+1}\; .
\end{equation} 
The operator $\mathbf{\Delta}_4$ appears instead for the first time at three loops
and contains quadrupole colour correlations among all four external legs.
It can also be expanded in $\as$ as
\begin{equation}
\mathbf{\Delta}_4 (\{p\}) = \sum_{L=3}^\infty \left(\frac{\as}{4 \pi}\right)^L \; \mathbf{\Delta}^{(L)}_4 (\{p\}) \; .  \label{delta4}
\end{equation}
In terms of these quantities one can organise the IR poles of the helicity amplitudes as 
\begin{align}
\mathbfcal{H}_{i,\:\text{fin}}^{(0)} &= \mathbfcal{H}_{i}^{(0)} \; ,  \label{tree}\\ 
\mathbfcal{H}_{i,\:\text{fin}}^{(1)} &= \mathbfcal{H}_{i,\: \text{ren}}^{(1)} - \mathbfcal{I}_1 \; \mathbfcal{H}_{i,\: \text{ren}}^{(0)} \; , \label{one}  \\ 
\mathbfcal{H}_{i,\:\text{fin}}^{(2)} &= \mathbfcal{H}_{i,\: \text{ren}}^{(2)} - \mathbfcal{I}_2\; \mathbfcal{H}_{i,\: \text{ren}}^{(0)} - \mathbfcal{I}_1\; \mathbfcal{H}_{i,\: \text{ren}}^{(1)} \; ,\label{two}  \\ 
\mathbfcal{H}_{i,\:\text{fin}}^{(3)} &= \mathbfcal{H}_{i,\: \text{ren}}^{(3)} - \mathbfcal{I}_3\; \mathbfcal{H}_{i,\: \text{ren}}^{(0)} - \mathbfcal{I}_2\; \mathbfcal{H}_{i,\: \text{ren}}^{(1)} - \mathbfcal{I}_1 \;\mathbfcal{H}_{i,\: \text{ren}}^{(2)} \; ,\label{three}  
\end{align}
where the IR subtraction operators read
\begin{align}
\mathbfcal{I}_{1}  &= \mathbfcal{Z}_1 \label{I1}\,,\\
\mathbfcal{I}_{2} &=  \mathbfcal{Z}_2 - \mathbfcal{Z}_1^2  \label{I2}\,, \\ 
 \mathbfcal{I}_{3} &= \mathbfcal{Z}_3    -   2\mathbfcal{Z}_1  \mathbfcal{Z}_2 + \mathbfcal{Z}_1^3 + \mathbf{\Delta}_4^{(3)}  \;,\label{I3}
\end{align}
with 
\begin{align}
 \mathbfcal{Z}_1 &=  \frac{\Gamma'_0}{4 \epsilon^2} + \frac{\mathbf{\Gamma}_0}{2 \epsilon} \,,   \label{Z1}\\
\mathbfcal{Z}_2 &=   \frac{{\Gamma_0'}^2}{32 \epsilon^4} + \frac{\Gamma'_0}{8 \epsilon^3} \left( \mathbf{\Gamma}_0 - \frac{3}{2} \beta_0  \right) +  \frac{\mathbf{\Gamma}_0}{8 \epsilon^2}(\mathbf{\Gamma}_0 - 2 \beta_0)  + \frac{\Gamma_1'}{16 \epsilon^2} + \frac{\mathbf{\Gamma}_1}{4 \epsilon}\,, \label{Z2}\\ 
\mathbfcal{Z}_3 &=  \frac{{\Gamma'_0}^3}{384 \epsilon^6}  + \frac{{\Gamma'_0}^2}{64 \epsilon^5}(\mathbf{\Gamma}_0 - 3 \beta_0) + \frac{\Gamma_0'}{32 \epsilon^4} \left( \mathbf{\Gamma}_0 - \frac{4}{3} \beta_0 \right) \left( \mathbf{\Gamma}_0 - \frac{11}{3} \beta_0 \right)  + \frac{\Gamma_0' \Gamma_1'}{64 \epsilon^4}  \nonumber\\
& \quad +\frac{\mathbf{\Gamma}_0}{48\epsilon^3}(\mathbf{\Gamma}_0 - 2 \beta_0)(\mathbf{\Gamma}_0 - 4 \beta_0) + \frac{\Gamma'_0}{16 \epsilon^3} \left( \mathbf{\Gamma}_1 - \frac{16}{9} \beta_1\right) + \frac{\Gamma_1'}{32 \epsilon^3} \left( \mathbf{\Gamma}_0 - \frac{20}{9} \beta_0 \right)  \nonumber\\ 
&\quad + \frac{\mathbf{\Gamma}_0 \mathbf{\Gamma}_1}{8 \epsilon^2} - \frac{\beta_0 \mathbf{\Gamma}_1 + \beta_1 \mathbf{\Gamma}_0}{6 \epsilon^2} + \frac{\Gamma_2'}{36 \epsilon^2 } + \frac{\mathbf{\Gamma}_2 + \mathbf{\Delta}_4^{(3)}}{6 \epsilon} \; ,\label{Z3}
\end{align}
together with the quadrupole contribution $\mathbf{\Delta}_4^{(3)}$, 
which is non-diagonal in colour space and reads explicitly
\begin{equation}\label{delta_4_main}
\mathbf{\Delta}_4^{(3)} = {\small
\begin{pmatrix}
 - 8 N_c \left[2 \left(D_2+2 D_1 \right) + 3 C \right] & 8 N_c^2 \left[2 D_2 - C\right]+32 \left[C-  D_1- D_2 \right] \\[8pt]
 8 N_c^2 \left[2 D_2- C\right]+32(C+ D_1) & 8 N_c \left[2 \left(D_2+2 D_1\right)-3 C\right] 
\end{pmatrix}
} \; .
\end{equation}
We stress here that, even if their $x$ dependence has been left implicit for clarity, $D_1$ and $D_2$ are non trivial functions of the kinematics.
More explicitly, the abbreviations in~\eqref{delta_4_main} read
\begin{align}
\label{C}
C &=  \zeta_5 + 2 \zeta_2 \zeta_3\,,
\\[2ex]
D_1 &= -2 \textit{G}_{1,4}-\textit{G}_{2,3}-\textit{G}_{3,2}+2 \textit{G}_{1,1,3}+2 \textit{G}_{1,2,2}-2 \textit{G}_{1,3,0}-\textit{G}_{2,2,0}-\textit{G}_{3,1,0} \nonumber\\
&\quad +2 \textit{G}_{1,1,2,0}-2 \textit{G}_{1,2,0,0}+ 2 \textit{G}_{1,2,1,0}+4 \textit{G}_{1,0,0,0,0}-2 \textit{G}_{1,1,0,0,0}+\frac{1}{2}\zeta_5  - 5 \zeta_2 \zeta_3  \nonumber \\
 &\quad  + \zeta_2 [5 \textit{G}_{3}+5 \textit{G}_{2,0}+2 \textit{G}_{1,0,0}-6 (\textit{G}_{1,2}+\textit{G}_{1,1,0})] + \zeta_3 (\textit{G}_{2}+2 \textit{G}_{1,0}-2 \textit{G}_{1,1}) \nonumber\\
&\quad - i \pi  [-\zeta_3 \textit{G}_{0}+\textit{G}_{2,2}+\textit{G}_{3,0}+\textit{G}_{3,1}+ \textit{G}_{2,0,0}+2 (\textit{G}_{1,3}-\textit{G}_{1,1,2}-\textit{G}_{1,2,1}-\textit{G}_{1,0,0,0})]\nonumber\\
&\quad + i \pi \zeta_2 (-\textit{G}_{2}+2 (\textit{G}_{1,1}+\textit{G}_{1,0}))- 11 i\pi \zeta_4 \, ,\label{D1}\\[10pt]
D_2 &=  2 \textit{G}_{2,3}+2 \textit{G}_{3,2}-\textit{G}_{1,1,3}-\textit{G}_{1,2,2}-2 \textit{G}_{2,1,2}+2 \textit{G}_{2,2,0}-2 \textit{G}_{2,2,1} \nonumber\\
&\quad +2 \textit{G}_{3,1,0}-2 \textit{G}_{3,1,1}-\textit{G}_{1,1,2,0}- \textit{G}_{1,2,1,0}-2 \textit{G}_{2,1,1,0}+4 \textit{G}_{2,1,1,1}-\zeta_5 +4 \zeta_2 \zeta_3 \nonumber \\
&\quad + \zeta_3 \textit{G}_{1,1}+\zeta_2 [-6 \textit{G}_{3}-6 \textit{G}_{2,0}+2 \textit{G}_{2,1}+5 (\textit{G}_{1,2}+\textit{G}_{1,1,0})] \nonumber\\
&\quad + i \pi  (\zeta_3 \textit{G}_{1}+2 \textit{G}_{3,0}-\textit{G}_{1,1,2}-\textit{G}_{1,2,0}-\textit{G}_{1,2,1}+2 \textit{G}_{2,0,0}-2 \textit{G}_{2,1,0} \nonumber\\
&\quad +2 \textit{G}_{2,1,1}-\textit{G}_{1,1,0,0})+ i \pi \zeta_2  (4 \textit{G}_{2}-\textit{G}_{1,1}) \, .  \label{D2}
\end{align}
where $D_1$ and $D_2$ are expressed in terms of generalized polylogarithms of argument $x$ with letters 0 and 1.
In eqs.~(\ref{D1},\ref{D2}) we have suppressed the $x$ dependence and used a compact index notation for the harmonic polylogarithms similar to \cite{Remiddi:1999ew,Maitre:2005uu},
\begin{align}
\label{compactindex}
G_{a_1,\dots,a_n,\footnotesize\underbrace{ 0,\dots,0}_{n_0}} &= G(\underbrace{0,\dots,0}_{|a_1|-1},\sgn(a_1),\dots,\underbrace{0,\dots,0}_{|a_n|-1},\sgn(a_n),\underbrace{0,\dots,0}_{n_0};x),
\end{align}
for $a_i \in \mathbbm{Z} \setminus\!\{0\}$.
The $G$ functions on the r.h.s.\ of eq.~\eqref{compactindex} are the generalized polylogarithms
\begin{align}
G(w_1,w_2,\dots,w_n; x)&= \int_0^x \frac{dt}{t-w_1} G(w_2,\dots,w_n; t) \quad \text{if at least one $w_i\neq 0$}, \\
G(\underbrace{0,\ldots,0}_{n}; x) &= \frac{1}{n!}\log^{n}(x),
\end{align}
and we use this notation also to present our explicit analytical results for the finite remainders.
We note here that both $D_1$ and $D_2$ are of uniform transcendental of weight five and are independent of the matter content of the theory, $i.e.$\ they do not depend
explicitly on $n_f$. 
More details on the derivation of these formulas are provided in Appendix~\ref{sec:appB}.

%% file: 6-results.tex

\section{Results} \label{results}

After the ultraviolet and infrared pole subtraction described in the previous section, we arrive at the main result of this paper, the fully analytical expressions for the finite remainders of the helicity amplitudes for process \eqref{s_channel}.
As previously mentioned,  the helicity amplitudes for other $2\to2$ quark processes with different initial states, including the equal-flavour case $q=Q$, can be obtained by a combination of analytical continuation and momenta renaming from the ones for our main process \eqref{s_channel}.  We  discuss this in more detail in Section \ref{extra_results}.
We provide all finite remainders  in electronic format as ancillary files attached
to the \texttt{arXiv} submission of this manuscript.

\subsection{Checks}
We have performed various checks on our results. First of all, we have verified that the IR poles of our scattering
amplitudes  follow the pattern predicted in refs~\cite{Becher:2009qa,Becher:2009cu,Almelid:2015jia} up to three loops.
We have then checked the finite part of our one loop amplitudes for all different partonic channels against the automated one-loop generator \texttt{OpenLoops}~\cite{Cascioli:2011va,Buccioni:2019sur}.
Finally, we have checked our one- and two-loop amplitudes through to order $\epsilon^4$ and $\epsilon^2$, respectively,
 against the results presented in refs~\cite{Glover:2004si,Ahmed:2019qtg}.

In order to successfully perform this check, one has to pay particular attention when comparing the amplitudes before IR subtraction, 
due to a subtlety in the  dimensional regularisation scheme 
used in refs~\cite{Glover:2004si,Ahmed:2019qtg}. This is due to the fact that the tensor structures
used to decompose the scattering amplitude in those references contain an explicit dependence on the dimensional-regulation
parameter $\epsilon$, even if the external states are taken to be in four dimensions, as in the 't Hooft-Veltman prescription.
While ignoring this dependence does not change the finite remainder of the scattering amplitudes after UV and IR poles have been subtracted,
it does change the bare results.
We illustrate this point explicitly for the one loop case. 
In refs \cite{Glover:2004si,Ahmed:2019qtg},  the following four tensors are used to decompose the amplitude at one loop
in CDR\footnote{Note that here even $\widetilde T_1$  and  $\widetilde T_2$ do not coincide with our definitions.}

\begin{align} 
\label{tensors_henn_D1} \widetilde T_1 &= {\bar u} (p_2)\: \slashed p_3 \: { u} (p_1) \times   {\bar u} (p_4) \: \slashed p_1 \:{ u}(p_3) \; , \\
\label{tensors_henn_D2} \widetilde T_2 &= {\bar u} (p_2)\: \gamma^\alpha \: { u} (p_1) \times {\bar u} (p_4) \: \gamma_\alpha \:{ u}(p_3) \; , \\
\label{tensors_henn_D3} \widetilde T_3 &= {\bar u} (p_2)\: \slashed p_3   \gamma^\mu   \gamma^\nu \: { u} (p_1) \times   {\bar u} (p_4) \:
 \slashed p_1 \gamma_\mu  \gamma_\nu  \:{ u}(p_3) \; , \\
\label{tensors_henn_D4} \widetilde T_4 &= {\bar u} (p_2)\: \gamma^\alpha \gamma^\mu  \gamma^\nu \: { u} (p_1) \times {\bar u} (p_4) \: \gamma_\alpha  \gamma_\mu   \gamma_\nu \:{ u}(p_3) \; , 
\end{align} 
where we stress that this decomposition is loop dependent and is only sufficient up to one-loop order, $L \leq 1$
\begin{equation}\label{decomp_tensors_henn}
\bar{\mathbfcal{A}}^{L \leq 1}  =   \widetilde{\mathbfcal{F}}_1 \; \widetilde{T}_1 + \widetilde{\mathbfcal{F}}_2 \; \widetilde{T}_2 + \widetilde{\mathbfcal{F}}_3 \; \widetilde{T}_3 + \widetilde{\mathbfcal{F}}_4 \; \widetilde{T}_4.
\end{equation}

Importantly, in the 't Hooft-Veltman scheme the vector indices of the $\gamma$ matrices in eqs.~\eqref{tensors_henn_D1}---\eqref{tensors_henn_D4} that are not 
explicitly contracted with four-dimensional vector fields 
are in general to be taken in 
$d$ dimensions. While this makes no difference for the 
first two tensors, the second two~(\ref{tensors_henn_D3},\ref{tensors_henn_D4})  
depend explicitly on $d$ and are responsible for an ambiguity 
in the way the helicity amplitudes are defined.
Let us consider for example the fourth tensors, eq.~\eqref{tensors_henn_D4}. 
If we consider the $\gamma$ matrices to carry $d$ dimensional vector indices we can
split the four dimensional part from the
$\epsilon$-dependent one by writing 
$\gamma^\mu_d = \gamma^\mu_4 + \gamma^{\mu}_{-2 \epsilon}$ and then use the equation 
\begin{equation}
    {\rm Tr}[\gamma_4^{\mu_1} \dots \gamma_4^{\mu_n} \gamma_{-2\epsilon}^{\nu_1} \dots \gamma_{-2\epsilon}^{\nu_m} ] = \frac{1}{4}     {\rm Tr}[\gamma_4^{\mu_1} \dots \gamma_4^{\mu_n} ]     {\rm Tr}[ \gamma_{-2\epsilon}^{\nu_1} \dots \gamma_{-2\epsilon}^{\nu_m} ] 
\end{equation}
as done in ref.~\cite{Cullen:2010jv} to extract the $(-2\epsilon)$-dimensional dependence of the $\gamma$-strings. All traces can then be evaluated as usual using the Clifford algebra relation $\{\gamma_\mu,\gamma_\nu\} = 2 g_{\mu \nu}$ for $d$-dimensional indices $\mu,\nu$. In this case, the dimensional splitting procedure simply amounts to $\epsilon$ dependent coefficients of the 4-dimensional $\gamma$-strings. For instance, taking the first string of $\widetilde{T}_4$ we find  
\begin{align}
    {\bar u} (p_2)\: \gamma^\alpha \gamma^\mu  \gamma^\nu \: { u} (p_1)& = g_{-2\epsilon}^{\mu \nu}  {\bar u} (p_2)\: \gamma_4^\alpha \: { u} (p_1) \nonumber \\
    &+ g_{-2\epsilon}^{\mu \alpha}  {\bar u} (p_2)\: \gamma_4^\nu \: { u} (p_1) \nonumber \\ 
    &-  g_{-2\epsilon}^{\nu \alpha}  {\bar u} (p_2)\: \gamma_4^\mu \: { u} (p_1),
\end{align}
where ${\rm Tr}[\gamma_{-2\epsilon}^\mu \gamma_{-2\epsilon}^\nu] = g_{-2\epsilon}^{\mu \nu} $ is the $(-2\epsilon)$-dimensional part of the metric tensor. Repeating the exercise for the other fermion string and then fixing the external helicities to $(+,-,+,-)$ 
we find
\begin{equation}
\widetilde T_4|_{(+,-,+,-)} = (32 - 12 \epsilon) \; [24]\langle 3 1 \rangle\,.
\end{equation}
Similarly, by repeating the same exercise for both helicities as we did 
in eqs.~\eqref{H_ij},  
but this time starting from the tensor decomposition in eq.~\eqref{decomp_tensors_henn},
we find
\begin{align}
\mathbfcal{H}_1 &=  -tu\: \left[ \widetilde{\mathbfcal{F}}_1 + 8 \widetilde{\mathbfcal{F}}_3 \right]
+2 t\left[ \widetilde{\mathbfcal{F}}_2 + 16 \widetilde{\mathbfcal{F}}_4 \right]  
+ 2 \epsilon \left[ t u\: \widetilde{\mathbfcal{F}}_3 - 6t \widetilde{\mathbfcal{F}}_4 \right] \: ,  \label{matrix_H1}\\
\mathbfcal{H}_2 &=  -tu \: \left[ \widetilde{\mathbfcal{F}}_1 + 4 \widetilde{\mathbfcal{F}}_3\right] 
- 2 u\left[ \widetilde{\mathbfcal{F}}_2 + 4 \widetilde{\mathbfcal{F}}_4 \right] 
+ 2\epsilon \left[ t u \:\widetilde{\mathbfcal{F}}_3 + 6 u\:\widetilde{\mathbfcal{F}}_4 \right] \; .\label{matrix_H2}
\end{align}
It is instructive to compare these formulas to the corresponding ones 
obtained in our approach, see eq.~\eqref{H1H2def}.
Our expressions for the helicity amplitudes, despite not displaying 
any explicit dependence on the parameter
$\epsilon$,  are exact in the 't Hooft-Veltman scheme. 
Notice, in particular, that the two tensors onto which we decompose the
amplitude,
${T}_1$ and ${T}_2$ in eq.~\eqref{tensors}, 
have been chosen such that it makes no practical difference
whether the $\gamma$ algebra to fix the helicity amplitudes
is performed $4$ or in $d=4 - 2 \epsilon$ dimensions. 
Upon substituting the form factors provided in refs~\cite{Glover:2004si,Ahmed:2019qtg} in eqs.~\eqref{matrix_H1} and \eqref{matrix_H2} (and in the corresponding
generalisations for the two-loop corrections),
we find perfect agreement up to weight six with 
our results for the \emph{bare helicity amplitudes}.

We stress, nevertheless, that the results for the bare helicity amplitudes as provided in refs~\cite{Glover:2004si,Ahmed:2019qtg} are obtained 
by setting $\epsilon=0$ in the coefficients of the form factors 
of \eqref{matrix_H1} and \eqref{matrix_H2} before substituting
the results for the form factors. This amounts to having assumed that the $\gamma$ matrices in eq.~\eqref{decomp_tensors_henn} are purely four-dimensional.  
This produces a difference for the bare amplitudes with respect to ours 
of order $\mathcal{O}(\epsilon)$ 
at one loop and $\mathcal{O}(1/\epsilon)$ at 
two loops.\footnote{Note that in this approach, one needs two more tensors 
at two loops, which depend quadratically on $\epsilon$.}  
However, it is easy to see that, as long as this choice is made 
consistently to all orders,  one obtains the same results for the 
finite remainders in $d=4$. 
In fact, one can imagine to first subtract UV and IR poles at the level of 
the individual form factors $\widetilde{\mathbfcal{F}}_j$ and, only afterwards,
substitute the finite form factors in eqs.~\eqref{matrix_H1} and \eqref{matrix_H2}, and fix $\epsilon = 0$. 
It is then obvious that the finite remainder cannot depend on 
the $\epsilon$-suppressed contributions in eqs.~\eqref{matrix_H1} 
and \eqref{matrix_H2}.
 We have verified the last statement directly, finding perfect agreement with refs~\cite{Glover:2004si,Ahmed:2019qtg} at the level
 of the finite remainders. 

\subsection{Numerical Evaluation}

\begin{figure}
\center
\includegraphics[width=1\textwidth]{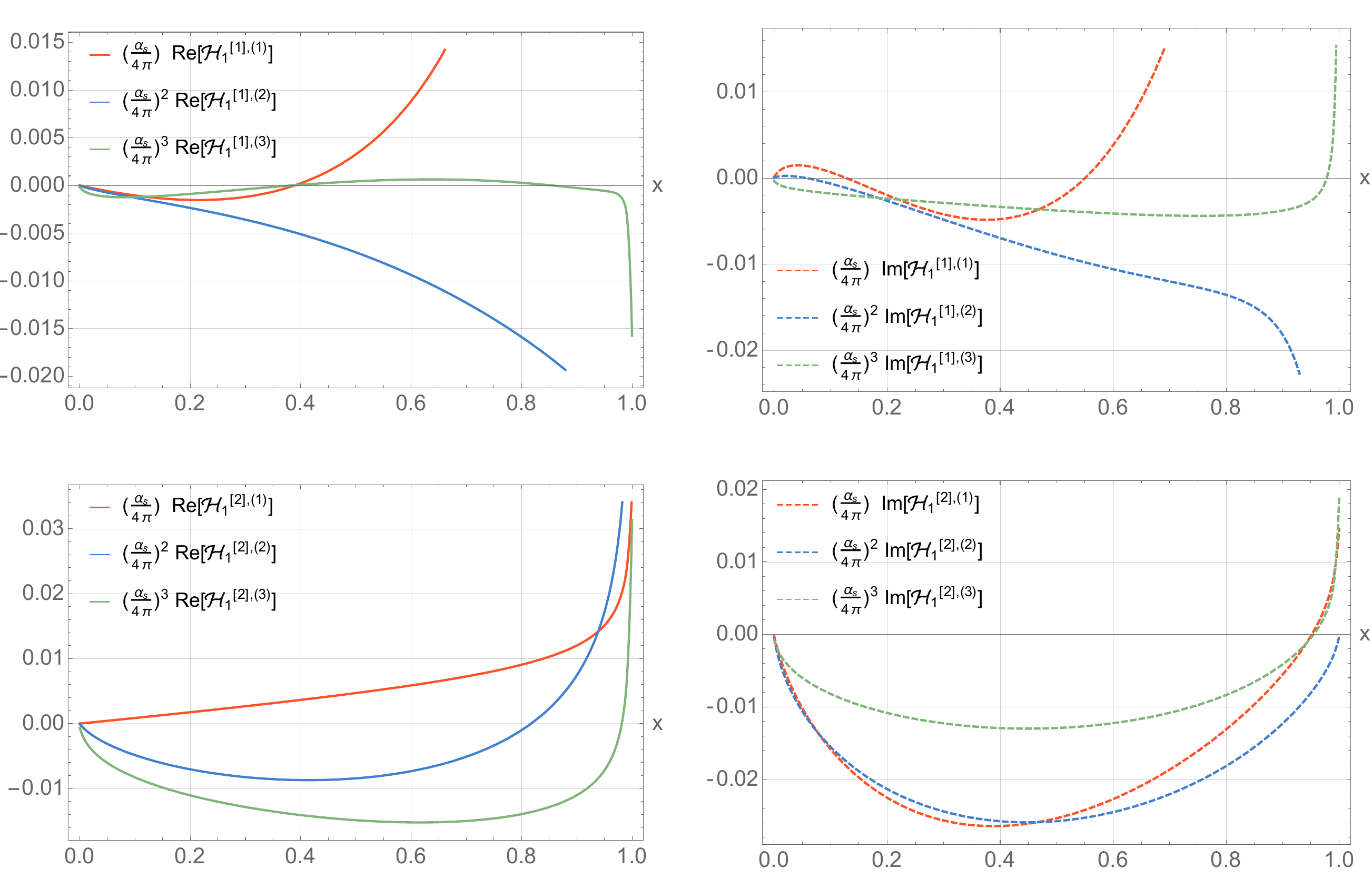}
\caption{Real (left) and imaginary (right) parts of the form factors $\mathbfcal{H}_{1,\text{fin}}^{[i],(L)}$ relevant for helicities $(+,-,+,-)$.   Colour components $[i]$ and number of loops $(L)$ are specified in the legends.}
\label{allH_1}
\end{figure} 

\begin{figure}
\center
\includegraphics[width=1\textwidth]{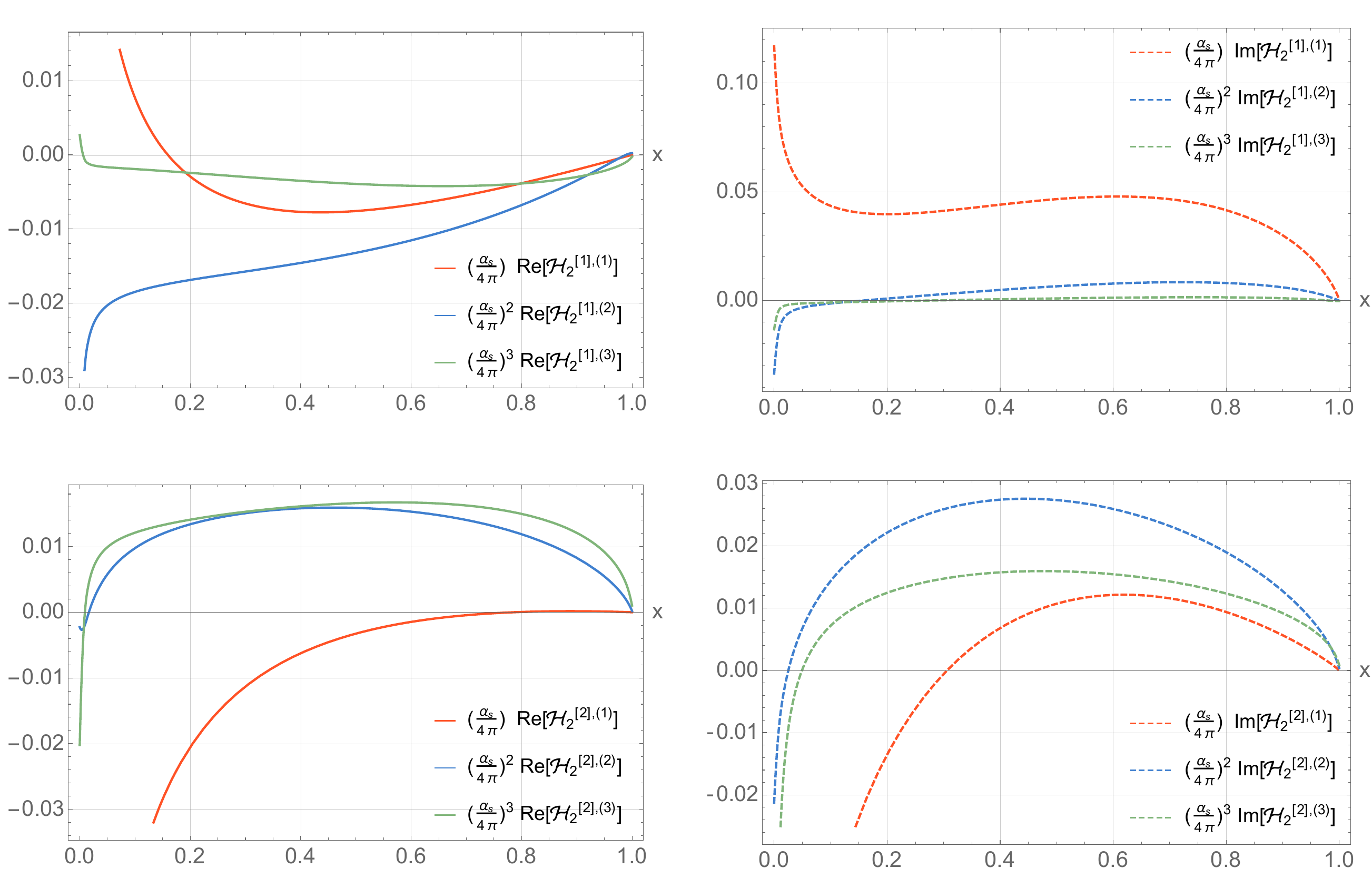}
\caption{Real (left) and imaginary (right) parts of the form factors $\mathbfcal{H}_{2,\text{fin}}^{[i],(L)}$ relevant for helicities $(+,-,-,+)$.   Colour components $[i]$ and number of loops $(L)$ are specified in the legends. }
\label{allH_2}
\end{figure} 
We present numerical results for the finite form factors defined in  \eqref{one}, \eqref{two} and \eqref{three} calculated in the physical region $0<x<1$ for the process in eq.~\eqref{s_channel}, $q\bar{q}\to Q\bar{Q}$. To evaluate our results numerically we made use of the \texttt{Mathematica} package \texttt{PolyLogTools} \cite{Duhr:2019tlz},
which in turn uses the \texttt{Ginac} library~\cite{Bauer:2000cp,cln,Vollinga:2004sn}.
For the various parameters we use the following values:
\begin{equation}
N_c = 3, \quad n_f = 5, \quad \as = 0.118, \quad  \mu^2 = s.
\end{equation}
We show results for $\mathbfcal{H}_{1,\text{fin}}$ corresponding to helicities $(+,-,+,-)$ in figure~\ref{allH_1} and for $\mathbfcal{H}_{2,\text{fin}}$ corresponding to helicities $(+,-,-,+)$ in figure~\ref{allH_2}.
In the figures, we present the two colour components of the form factors individually, where we recall that our colour decomposition reads: 
\begin{equation}
{\mathbfcal{H}}_{i,\text{fin}}^{(L)}  
= 
\begin{pmatrix}
{\mathcal{H}}_{i,\text{fin}}^{[1],(L)}  \\
{\mathcal{H}}_{i,\text{fin}}^{[2],(L)} 
\end{pmatrix} \quad i=1,2 \; .
\end{equation}
Here, the colour index $[1]$ is related to the colour structure  ${\delta}_{ i_1 i_4} {\delta}_{i_2 i_3}$ while the index $[2]$ refers to the coefficient of ${\delta}_{ i_1 i_2} {\delta}_{ i_3 i_4}$.
Lastly, the index $(L)$ refers to the number of loops of the corresponding amplitude.

%% file: 7-other_processes.tex

\section{Crossed Channels and Equal Flavour Amplitudes} \label{extra_results}
In the previous sections, we presented the computation of the helicity amplitudes 
for the scattering of four quarks of different flavour as in eq.~\eqref{s_channel}.  
We discuss here how to use this result to derive the helicity amplitudes for all other 2-to-2 quark scattering processes, both for different and equal flavours.
We start by listing all of these processes.
\begin{itemize}
\item Different flavour quarks:
 \begin{align}
 & { q}(p_1) \;+ \;\bar { q}(p_2)  \; \longrightarrow \;  { \bar Q}(-p_3) \;+ \; {   Q}(-p_4) \; ,    \label{s_channel_new} \\[8pt]
&  { q}(p_1) \;+ \; { Q}(p_2)  \; \longrightarrow \;  {  q}(-p_3) \;+ \; {   Q}(-p_4) \; ,  \label{t_channel}   \\[8pt]
& { q}(p_1) \;+ \;\bar { Q}(p_2)  \; \longrightarrow \;  { \bar Q}(-p_3) \;+ \;{ q}(-p_4)  \; ,\label{u_channel} \\[8pt]
&  { \bar q}(p_1) \;+ \; { \bar Q}(p_2)  \; \longrightarrow \;  { \bar q}(-p_3) \;+ \; {  \bar  Q}(-p_4) \; .  \label{anti_t_channel} 
 \end{align}
 \item Equal flavour quarks:
\begin{align}
 & { q}(p_1) \;+ \;\bar { q}(p_2)  \; \longrightarrow \;  { \bar q}(-p_3) \;+ \; {   q}(-p_4) \; ,    \label{s_channel_equal} \\[8pt]
 & { q}(p_1) \;+ \; { q}(p_2)  \; \longrightarrow \;  {  q}(-p_3) \;+ \; {  q}(-p_4) \; ,  \label{t_channel_equal}  \\[8pt]
  & { \bar q}(p_1) \;+ \; { \bar q}(p_2)  \; \longrightarrow \;  { \bar  q}(-p_3) \;+ \; { \bar  q}(-p_4) \; .  \label{anti_t_channel_equal}  
\end{align}
\end{itemize}
Here, the first process \eqref{s_channel_new} is the one we have already computed,  
$i.e.$\ process \eqref{s_channel}, see eqs.~\eqref{H_ij} and~\eqref{H1H2def}.

In the following we will consider only two helicity amplitudes for each process, since,  as already discussed,  the other two can be obtained by a parity transformation\footnote{For these processes,
a parity transformation only
acts on the external spinors by the following transformation: $ \langle ij \rangle  \leftrightarrow [ji]$.} which flips the signs of helicities.
We start with the case $q \neq Q$.  From the helicity amplitudes \eqref{H_ij},  one can find those for  \eqref{t_channel} and \eqref{u_channel}  by performing appropriate crossings of particles from the initial to the final state.  This can be achieved by analytically continuing the results for \eqref{s_channel_new} to the appropriate kinematical region (see figure~\ref{KinematicsPath} for reference) and then renaming the momenta of the particles involved in the crossing to match the definitions of the processes above.  Amplitudes for the process \eqref{anti_t_channel} are then simply found by acting with a charge conjugation transformation\footnote{See the formulas below for the explicit action of charge conjugation on the helicity amplitudes. } on \eqref{t_channel}.
Details on how to perform the analytic continuation of the transcendental functions appearing in the scattering amplitudes are described for example in \cite{Anastasiou:2000mf}.  However,  for convenience of the reader,  we outline the method in Appendix \ref{sec:appC}.  
In the following we will abandon the colour-vector notation for helicity amplitudes as the crossing of particles also changes the indices of the colour basis \eqref{colour_structures}. Instead, we adopt the more conventional notation
\begin{equation}
\mathbfcal{H}_i =
\begin{pmatrix}
{\mathcal{H}}_i^{[1]}  \\
{\mathcal{H}}_i^{[2]} 
\end{pmatrix} 
\longrightarrow  \mathcal{H}_i = \mathcal{H}_i^{[1]} \mathcal{C}_1 +  \mathcal{H}_i^{[2]} \mathcal{C}_2.
\end{equation}
This way the crossing acts on the colour structure of the amplitude by simply exchanging indices in $\mathcal{C}_1$ and $\mathcal{C}_2$.
With this notation and following the procedure described above, we find\footnote{We recall that the notation $\bar{\mathcal{A}}$ is defined in eq.~\eqref{definition_A_bar}.}  for processes \eqref{t_channel} and \eqref{anti_t_channel}
\begin{align}
\bar{\mathcal{A}}_{++--}^{\scriptstyle qQ\rightarrow qQ} &=\bar{\mathcal{A}}_{++--}^{\scriptstyle \bar q\bar Q\rightarrow \bar q\bar Q} = \mathcal{H}_1|_{p_2 \leftrightarrow p_3}  \;  \frac{ \langle 12 \rangle}{\langle 34 \rangle},   \\[8pt]
\bar{\mathcal{A}}_{+--+}^{\scriptstyle qQ\rightarrow qQ} &= \bar{\mathcal{A}}_{+--+}^{\scriptstyle \bar q\bar Q\rightarrow \bar q\bar Q} =\mathcal{H}_2|_{p_2 \leftrightarrow p_3}  \; \frac{ \langle 14 \rangle}{\langle 32 \rangle}\,, \label{anti_t_channel_2}
\end{align}
and for process \eqref{u_channel} 
\begin{align}
\bar{\mathcal{A}}_{+-+-}^{\scriptstyle q\bar Q\rightarrow \bar Q q } &= \mathcal{H}_1|_{p_2 \leftrightarrow p_4}  \;  \frac{ \langle 13 \rangle}{\langle 42 \rangle} ,   \\ \bar{\mathcal{A}}_{++--}^{\scriptstyle q \bar Q\rightarrow \bar Q q } &= \mathcal{H}_2|_{p_2 \leftrightarrow p_4}  \;  \frac{ \langle 12 \rangle}{\langle 43 \rangle} \;.
\end{align}
We now turn to the processes with $q=Q$. 
To obtain the amplitude for \eqref{s_channel_equal}, we can use the fact that the Feynman diagrams contributing to this process are exactly the sum of the ones of processes \eqref{s_channel_new} and \eqref{u_channel}.  
Accounting for a relative minus sign between the two contributions to the amplitude due to the exchange of two identical fermions, we write, similarly to \cite{Glover:2004si},
\begin{equation}\label{equal_quarks_relation}
\bar{\mathcal{A}}_{\lambda_1\lambda_2\lambda_3\lambda_4}^{\scriptstyle q\bar q \rightarrow \bar qq} = \bar{\mathcal{A}}_{\lambda_1\lambda_2\lambda_3\lambda_4}^{\scriptstyle q\bar q \rightarrow \bar QQ}  - \bar{\mathcal{A}}_{\lambda_1\lambda_2\lambda_3\lambda_4}^{\scriptstyle q\bar Q \rightarrow \bar Q q}\,.
\end{equation}
For the two independent helicity choices we obtain
\begin{align}
&\bar{\mathcal{A}}_{+-+-}^{\scriptstyle q\bar q \rightarrow \bar  q q} = \Big[ \mathcal{H}_1 - \mathcal{H}_1|_{p_2 \leftrightarrow p_4} \Big] \; \frac{ \langle 13 \rangle}{\langle 24 \rangle},
 \label{s_equal_quarks_1}\\[8pt]
&\bar{\mathcal{A}}_{+--+}^{\scriptstyle q\bar q \rightarrow \bar  qq } = \mathcal{H}_2 \; \frac{ \langle 14 \rangle}{\langle 23 \rangle} ,\label{s_equal_quarks_2}
\end{align}
where in the second equation we used $ \bar{\mathcal{A}}_{+--+}^{\scriptstyle q\bar Q \rightarrow \bar Q q}  = 0$.
In the same way, we can find the amplitudes for processes \eqref{t_channel_equal} and \eqref{anti_t_channel_equal} as a sum of those for process \eqref{t_channel} and the ones obtained by crossing $p_3 \leftrightarrow p_4$ in \eqref{t_channel}. 
Explicitly, we obtain
\begin{equation}\label{equal_quarks_relation_2}
\bar{\mathcal{A}}_{\lambda_1\lambda_2\lambda_3\lambda_4}^{\scriptstyle q q \rightarrow q q} = \bar{\mathcal{A}}_{\lambda_1\lambda_2\lambda_3\lambda_4}^{\scriptstyle \bar q \bar q \rightarrow \bar q \bar q} = \bar{\mathcal{A}}_{\lambda_1\lambda_2\lambda_3\lambda_4}^{\scriptstyle qQ\rightarrow qQ}  - \bar{\mathcal{A}} ^{\scriptstyle qQ\rightarrow qQ}_{{\lambda_1\lambda_2\lambda_4\lambda_3}}\big|_{p_3 \leftrightarrow p_4},
\end{equation}
where we stress the exchange of helicity labels in the second term of the r.h.s.
This implies
\begin{align}
\bar{\mathcal{A}}_{++--}^{\scriptstyle q q \rightarrow q q} &=\bar{\mathcal{A}}_{++--}^{\scriptstyle \bar q \bar q \rightarrow \bar q \bar q} = \Big[ \mathcal{H}_1|_{p_2 \leftrightarrow p_3}  + \mathcal{H}_1|_{p_2 \rightarrow p_4 \rightarrow p_3 \rightarrow p_2}   \Big]  \; \frac{ \langle 12 \rangle}{\langle 34 \rangle}  \;,  \label{t_equal_quarks_1}\\[8pt]
\bar{\mathcal{A}}_{+--+}^{\scriptstyle qq \rightarrow q q} &=\bar{\mathcal{A}}_{+--+}^{\scriptstyle \bar q\bar q \rightarrow \bar q \bar q} = \mathcal{H}_2|_{p_2 \leftrightarrow p_3} \; \frac{ \langle 14 \rangle}{\langle 32 \rangle} , \label{t_equal_quarks_2}
\end{align}
where we have used that $\bar{\mathcal{A}}^{\scriptstyle qQ\rightarrow qQ}_{{+-+-}}=0$.\\

We point out that the formulas of this section are valid for the bare amplitudes as well as for the UV renormalised and the finite remainders after IR subtraction.
We thus arrive at the following representation for the finite remainders of the full helicity amplitudes:
\begin{align}
    \mathcal{A}^{\text{process}}_{\lambda_1\lambda_2\lambda_3\lambda_4\,\text{fin}}
    &= 4 \pi \alpha_s \Phi^{\text{process}}_{\lambda_1\lambda_2\lambda_3\lambda_4}
    \sum_{L=0}^3 \left(\frac{\alpha_s}{4\pi}\right)^L \mathcal{H}^{\text{process}\;(L)}_{\lambda_1\lambda_2\lambda_3\lambda_4,\,\text{fin}} 
    + \mathcal{O}(\alpha_s^5)\,
\end{align}
where $\Phi^{\text{process}}_{\lambda_1\lambda_2\lambda_3\lambda_4}$ are the corresponding spinor phases, see e.g.\ eq.~\eqref{H_ij}.
We list the spinor phases and provide explicit analytical results for the form factors $\mathcal{H}^{\text{process}\;(L)}_{\lambda_1\lambda_2\lambda_3\lambda_4,\,\text{fin}}$ in the ancillary files.

%% file: 8-conclusions.tex

\section{Conclusions}
\label{conclusions}
In this paper we presented the analytic calculation of the three-loop
helicity amplitudes for the process $q \bar q
\rightarrow Q \bar{Q}$ and all other related $2 \to 2$ quark
scattering processes in massless QCD.  Our results were obtained
through a tensor decomposition of the scattering amplitude in the 't
Hooft Veltman scheme, which allowed us to work with a minimal number
of independent basis tensors, exactly matching the number of
independent helicity amplitudes.  This method is particularly convenient when
applied to processes with multiple external fermion lines, where the
number of basis tensors in conventional dimensional regularisation
is known to increase with the number of
loops.
We employed modern integration-by-parts reductions based on finite field
arithmetic and syzygy techniques to perform the demanding mapping
of the three-loop integrals to a basis of master integrals.
For the master integrals, analytical solutions in terms of harmonic polylogarithms
were already available~\cite{Henn:2020lye}. 
We point out that the finite remainders
at three loops are remarkably compact with file sizes 
of $\sim 1$ MB for each partonic channel. 
We provide the analytical results as ancillary files attached to the 
\texttt{arXiv} submission of this manuscript.

This is the first three-loop calculation in full QCD for a process
that involves the scattering of four coloured particles.  In
particular, our results confirm, for the first time, the structure of the colour
quadrupole contribution to the IR poles of QCD scattering
amplitudes with non-trivial colour flow.  We performed various checks
on our results, both comparing them with previous analytical
calculations at one and two loops, and numerically with
\texttt{OpenLoops 2}.  The first test highlighted a
subtlety in the definition of the \emph{bare} helicity amplitudes 
in 't Hooft-Veltman scheme for processes
which involve the scattering of more than one pair of fermions.
In this case, using the standard projectors defined in
conventional dimensional regularisation, different results 
can be obtained depending on whether the 
$\gamma$ algebra to fix the helicity amplitudes is performed 
in $4$ or in $d=4-2 \epsilon$ dimensions. 
While no ambiguity arises at the level of finite remainders in four dimensions,
extra care should be paid when comparing divergent results.

The four-quark scattering processes presented here are arguably the least
complex among the ones involving four coloured particles. However, the
calculation of the three-loop amplitudes for $q \bar q \rightarrow gg$, $gg \rightarrow gg$ and the processes related by
crossings require no new concepts and can be performed following the steps described in this paper.

%% file: 9-appA.tex

\section{Details on the IR Structure}
\label{sec:appB}
In this appendix we provide details on the calculation of the IR subtraction operators given in eqs. \eqref{I1}, \eqref{I2} and \eqref{I3}.  We recall that one can write the finite remainder of the scattering amplitude by means of a multiplicative colour-space operator $\mathbfcal{Z}$ 
\begin{equation}\label{Zeta_A_B}
\mathbfcal{A}_\text{fin} (\{p\},\mu)= \lim_{\epsilon \rightarrow 0} \mathbfcal{Z}^{-1}(\epsilon,\{p\},\mu) \; \mathbfcal{A}_{\text{ren}}(\epsilon,\{p\}) \; ,
\end{equation}
which we can apply similarly to the form factors in eq.~\eqref{hel_ampls_ren}: 
\begin{equation}\label{Zeta_H_B}
\mathbfcal{H}_{i,\:\text{fin}} (\epsilon,\{p\})= \lim_{\epsilon \rightarrow 0} \mathbfcal{Z}^{-1}(\epsilon,\{p\},\mu) \; \mathbfcal{H}_{i,\:\text{ren}}(\epsilon,\{p\}) \; .
\end{equation}

For the case of four-quark scattering considered in this paper, the
$\mathbfcal{Z}$ colour operator can be represented as a 2-by-2 matrix,
which mixes the colour structures defined in
eq.~\eqref{colour_structures}.
It is possible to write the $\mathbfcal Z$ operator
as~\cite{Becher:2009qa,Becher:2009cu}
\begin{equation}\label{exponentiation_B}
\mathbfcal{Z} (\epsilon,\{p\},\mu) = \mathbb{P} \exp \left[
  \int_\mu^\infty \frac{\mathrm{d} \mu'}{\mu'}
  \mathbf{\Gamma}(\{p\},\mu')\right] \; ,
\end{equation}
where $\mathbb{P}$ is the \textit{path-ordering} symbol, {\it i.e.}\
operators are ordered from left to right with decreasing values of $\mu'$.
To proceed, it is  useful to decompose the 
anomalous dimension operator according to
\begin{equation}\label{anomalous_operator_B}
\mathbf{\Gamma}(\{p\},\mu) =  \mathbf{\Gamma}_{\text{dipole}}(\{p\},\mu)  + \mathbf{\Delta}_4 (\{p\}) 
\end{equation}
into a dipole and a quadrupole contribution.

The dipole matrix was given in eq.~\eqref{dipole}. We repeat its form here for convenience
\begin{equation}\label{dipole_B}
\mathbf{\Gamma}_{\text{dipole}}(\{p\},\mu)  =  \sum_{1\leq i < j \leq 4} \mathbf{T}^a_i \; \mathbf{T}^a_j  \; \gamma^\text{cusp}(\as) \; \log\left(\frac{\mu^2}{-s_{ij}-i\delta}\right)  \; + \; \sum_{i=1}^4 \; \gamma^i(\as) \; ,
\end{equation}
where $\as=\as(\mu)$, $\gamma^\text{cusp}(\as)$ is the cusp anomalous
dimension and $\gamma^i$ the anomalous dimension of the $i$-th
external particle, which only depends on whether $i$ is a quark or a
gluon.  The perturbative expansion for these quantities are given up
to order $\mathcal{O}(\as^3)$ in Appendix \ref{sec:appA}.
In eq.~\eqref{dipole_B}, the $\mathbf T^a_i$ colour operators are
related to the
$SU(N_c)$ generator associated with the external particle $i$.
Specifically, following the conventions of \cite{Catani:1996vz} they are defined as
\begin{alignat}{2}
(\mathbf{T}^a_i)_{b_i c_i} &=    \;- i {f ^a}_{b_i c_i} &~&\text{for a gluon},\\
(\mathbf{T}^a_i)_{i_i j_i} &=  \; + T^a_{i_i j_i} &&\text{for a final(initial) state quark(anti-quark)},\\
(\mathbf{T}^a_i)_{i_i j_i} &=  \; - T^a_{j_i i_i} &&\text{for a initial(final) state quark(anti-quark)}.
\end{alignat}
We now consider how \eqref{dipole_B} can be written for our process \eqref{s_channel} as a matrix in colour space. The operator $\mathbf{T}^a_1 \; \mathbf{T}^a_2$ acts on our colour space basis tensors as follows 
\begin{align}
 \left( \mathbf{T}^a_1 \; \mathbf{T}^a_2 \;  \mathcal{C}_1 \right)_{i_1 i_2 i_3 i_4} &= - T^a_{j_1 i_1} T^a_{i_2 j_2} \; (\delta_{ j_1 i_4} \delta_{j_2 i_3})=- \; \frac{1}{2} \left( \delta_{ i_1 i_2} \delta_{i_3 i_4 } - \frac{1}{N_c} \delta_{ i_1 i_4} \delta_{i_2 i_3}  \right), \\[10pt]
\left( \mathbf{T}^a_1 \; \mathbf{T}^a_2 \; \mathcal{C}_2  \right)_{i_1 i_2 i_3 i_4} &= - T^a_{j_1 i_1} T^a_{i_2 j_2} \; (\delta_{ j_1 j_2} \delta_{i_3 i_4 })= -C_F  \;  \delta_{ i_1 i_2} \delta_{i_3 i_4 }.
\end{align}
We change to component notation with respect to our $\mathcal{C}_i$ basis by identifying
\begin{equation}\label{ref1_start}
v_1 \;  \delta_{ i_1 i_4} \delta_{i_2 i_3} + v_2 \; \delta_{ i_1 i_2} \delta_{i_3 i_4 } \rightarrow \begin{pmatrix}
v_1 \\
v_2
\end{pmatrix}
\end{equation}
and obtain the matrix representation
\begin{equation}
\mathbf{T}^a_1 \; \mathbf{T}^a_2 = \begin{pmatrix}
\frac{1}{2N_c} & 0 \\
-\frac{1}{2} & -C_F 
\end{pmatrix} \; .
\end{equation}
Similarly, we find for the first sum in eq.~\eqref{dipole_B} the matrix representation
\begin{equation}\label{dipole_matrix_form_B}
  \sum_{1\leq i <  j \leq 4} \mathbf{T}^a_i \; \mathbf{T}^a_j
  \; \gamma^\text{cusp}(\as)\;  \log\left(\frac{\mu^2}{-s_{ij}-i\delta}\right) = \gamma^\text{cusp} (\as) \begin{pmatrix}
A(s,u) & B(s,u) \\[8pt]
B(u,s) & A(u,s)
\end{pmatrix} \; ,
\end{equation}  
with 
\begin{align}
A(s,u)&=-2\left\{C_F \log\left(\frac{\mu^2}{-u-i\delta}\right)+\frac{1}{2N_c}\left[ \log\left(\frac{\mu^2}{-t-i\delta}\right) - \log\left(\frac{\mu^2}{-s-i\delta}\right) \right] \right\} \: , \label{A_B}\\[10pt]
B(s,u)&= -\left[ \log\left(\frac{\mu^2}{-u-i\delta}\right) - \log\left(\frac{\mu^2}{-t-i\delta}\right) \right] \; , \label{ref1_end}
\end{align}
where $t=-s-u$ is implied.
In these equations, a small positive imaginary part is associated with the Mandelstam invariants to fix branch cut ambiguities.

\begin{figure}
\centering
\subfigure[]{\includegraphics[width=0.35\textwidth]{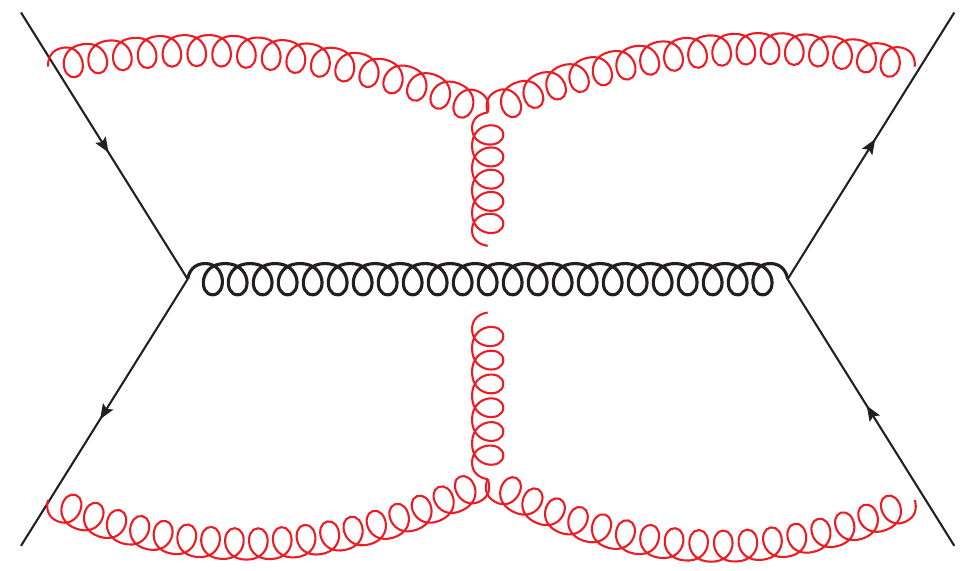}}\label{fig:2a}
\subfigure[] {\includegraphics[width=0.35\textwidth]{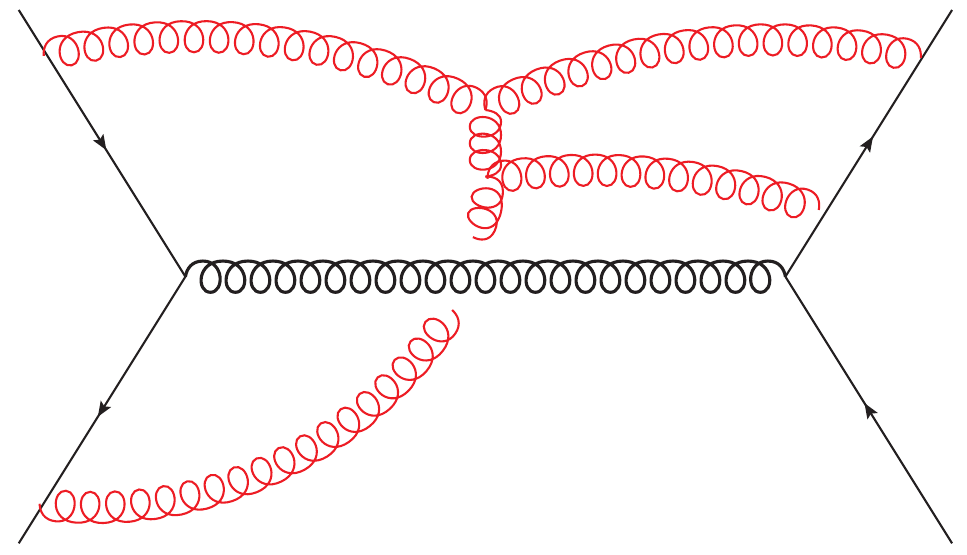}}\label{fig:2b}
\caption{Diagrams contributing to the quadrupole IR divergence, reinterpreted as tree level diagrams plus virtual gluons. Diagrams (a) and (b) give a contribution to the colour structures in the first and second line of eq.~\eqref{eq:quadrupole}, respectively.} \label{diagrams_red}
\end{figure}

The dipole anomalous matrix that we have just discussed is sufficient
to reproduce the IR poles of our scattering amplitude up to two
loops.  Starting at three loops, however, new four-parton colour
correlations contained in the matrix $\mathbf{\Delta}_4$ spoil the
simple dipole picture.  To compute $\mathbf{\Delta}_4$ explicitly,
we start by writing
\begin{equation}
\mathbf{\Delta}_4 (\{p\}) = \sum_{L=3}^\infty \left(\frac{\as}{4 \pi}\right)^L \; \mathbf{\Delta}^{(L)}_4 (\{p\}) \; ,  \label{delta4_B}
\end{equation}
such that, at the three-loop level, we are only interested in the first term of the expansion.
By analytically continuing eq.~(7) of Ref.~\cite{Almelid:2015jia} to the kinematical region with $s>0$, $t,u < 0$,  we find
\begin{align} \label{eq:quadrupole}
\mathbf{\Delta}^{(3)}_4 &= \;  128 \; f_{abe} \: f_{cde} \;  \left[ \mathbf{T}^a_1\:\mathbf{T}^c_2\:\mathbf{T}^b_3\:\mathbf{T}^d_4\: D_1(x) - \mathbf{T}^a_4\:\mathbf{T}^b_1\:\mathbf{T}^c_2\:\mathbf{T}^d_3\: D_2(x) \right] \nonumber\\
&\quad - 16 \; C \; f_{abe} \: f_{cde}\;  \sum_{i=1}^4 \sum_{\substack{1\leq j < k \leq4 \\ j,k\neq i}}  \left\{ \mathbf{T}^a_i,\mathbf{T}^d_i \right\} \; \mathbf{T}^b_j \; \mathbf{T}^c_k \; ,
\end{align}
with $x=-t/s$ given in eq.~\eqref{variables}, the constant $C$ in eq.~\eqref{C}, and the functions $D_1$, $D_2$ in eqs.~\eqref{D1}, \eqref{D2}. We note that all of the four partons of the kinematically dependent parts of eq.~\eqref{eq:quadrupole} are correlated through colour, while the kinematically independent part has a three parton colour correlation.
We also note that, at this level, the quadrupole corrections to the {\it{anomalous dimension}} matrix can be generated solely by gluon interactions, and hence it is expected to be universal for all gauge theories involving the gluon in the particle content. Two of the representative diagrams that give rise to the quadrupole corrections are those in panels (e) and (f) of figure~\ref{diagrams}. 
We can highlight contributions to the quadrupole divergence, and in particular to the colour structures in the first and second line of eq.~\eqref{eq:quadrupole}, by drawing the representative diagrams in figure~\ref{diagrams_red},
where the black lines represent a tree level diagram and the red lines the gluons responsible for the soft quadrupole corrections.
We stress here that in reference~\cite{Almelid:2015jia}
the authors give results for $\mathbf{\Delta}^{(3)}_4$ prior to imposing momentum conservation.
In this formulation, there exists a kinematical region where $\mathbf{\Delta}^{(3)}_4$ 
is purely real, which corresponds to $s_{ij}<0$ for all $i,j$.  
Consequently, we first analytically continue to the region in which $s= s_{12} > 0$ and then impose momentum conservation.

Just like for the dipole case
it is useful to represent the quadrupole operator as a matrix acting 
in colour space.  This is immediate to do for the first line of $\mathbf{\Delta}_4^{(3)}$ in eq.~\eqref{eq:quadrupole},
since each operator $\mathbf{T}$ that contributes to it acts on a different parton 
and thus commutes with all  other colour operators.
More care is required instead to manipulate the second line of 
eq.~\eqref{eq:quadrupole} since it contains operators acting on the same particle: 
this means that the ordering of the $\mathbf{T}_i$ is relevant.  
Here, the computation can be performed following again the procedure employed for the dipole case, see 
eqs.~\eqref{ref1_start}---\eqref{ref1_end}. The
result is presented in eq.~\eqref{delta_4_main}.

With the dipole and quadrupole contributions at hand,
it is easy to compute the $\mathbfcal{Z}$ opperator using eq.~\eqref{exponentiation_B} and to perform the exponentiation as a series expansion in $\as$.
To do so, we follow ref.~\cite{Becher:2009qa} and
introduce the short-hand
\begin{equation}
  \Gamma' = \frac{\partial}{\partial \log \mu} \mathbf{\Gamma}_{\text{dipole}} =
  \sum_{1\leq i < j \leq 4} \mathbf{T}^a_i \; \mathbf{T}^a_j\,
  \gamma^\text{cusp}(\as) = - 4 C_F\, \gamma^\text{cusp}(\as),
\end{equation}
and the $\as$ expansions
\begin{equation}
\mathbf{\Gamma}_{\text{dipole}} = \sum_{n=0}^\infty \mathbf{\Gamma}_n \;  \left( \frac{\as}{4 \pi} \right)^{n+1}, \quad  \quad \Gamma '=\sum_{n=0}^\infty \Gamma'_n \;  \left( \frac{\as}{4 \pi} \right)^{n+1}\; .
\end{equation}
In terms of these quantities, it is easy to find
\begin{align}\label{logZ_B}
&\log \mathbfcal{Z} = \frac{\as}{4 \pi}  \left[ \frac{\Gamma'_0}{4 \epsilon^2} + \frac{\mathbf{\Gamma}_0}{2 \epsilon}\right] \nonumber\\[8pt]
& + \left( \frac{\as}{4 \pi} \right)^2  \left[- \frac{3 \beta_0 \Gamma'_0}{16 \epsilon^3} + \frac{\Gamma'_1 - 4 \beta_0 \mathbf{\Gamma}_0}{16 \epsilon^2} + \frac{\mathbf{\Gamma}_1}{4 \epsilon}  \right] \\[8pt]
&+\left(\frac{\as}{4 \pi}\right)^3 \left[ \frac{11 \beta_0^2 \Gamma_0'}{72 \epsilon^4}  - \frac{5 \beta_0  \Gamma_1' + 8 \beta_1 \Gamma_0' - 12 \beta_0^2 \mathbf{\Gamma}_0}{72 \epsilon^3} + \frac{\Gamma_2' - 6 \beta_0 \mathbf{\Gamma}_1 - 6 \beta_1 \mathbf{\Gamma}_0}{36 \epsilon^2}  + \frac{\mathbf{\Gamma}_2 + \mathbf{\Delta}^{(3)}_4}{6 \epsilon}\right] \; .\nonumber
\end{align} 
From this result, it is straightforward to obtain the perturbative
expansion of $\mathbfcal Z$,
\begin{equation}
\mathbfcal{Z} = 1 + \left( \frac{\as}{4\pi} \right)
\mathbfcal{Z}_1+ \left( \frac{\as}{4\pi} \right)^2 \mathbfcal{Z}_2 +
\left( \frac{\as}{4\pi} \right)^3 \mathbfcal{Z}_3 + \mathcal O(\as^4)\;,
\end{equation}
and of its inverse $\mathbfcal Z^{-1}$,
\begin{equation}\label{Zinverse_B}
\mathbfcal{Z}^{-1} = 1 - \left( \frac{\as}{4\pi}  \right) \mathbfcal{I}_1- \left( \frac{\as}{4\pi}  \right)^2 \mathbfcal{I}_2 - \left( \frac{\as}{4\pi}  \right)^3 \mathbfcal{I}_3 + \mathcal O(\as^4)\;.
\end{equation}
The explicit form of the $\mathbfcal Z_i$ and $\mathbfcal I_1$ coefficients
are presented in eqs.~(\ref{I1}-\ref{Z3}).

%% file: 10-appB.tex

\section{Anomalous Dimension Coefficients}
\label{sec:appA}
In this appendix, we give the universal cusp anomalous dimension $\gamma^\text{cusp}$ and the quark anomalous dimension $\gamma^q = \gamma^{\bar q}$ up to $\mathcal{O}(\as^3)$ relevant for the three loop calculation of this paper.
We define the $\as$ expansion coefficients according to
\begin{equation}
\gamma^\text{cusp}=\sum_{n=0}^\infty \gamma^\text{cusp}_n \;  \left( \frac{\as}{4 \pi} \right)^{n+1}\; , \quad \quad \gamma^{q}=\sum_{n=0}^\infty \gamma^{q}_n \;  \left( \frac{\as}{4 \pi} \right)^{n+1}\; .
\end{equation}
We then have~\cite{Moch:2004pa,Vogt:2004mw}
\begin{eqnarray}
   \gamma_0^\text{cusp} &=& 4 \,, \nonumber\\
   \gamma_1^\text{cusp} &=& \left( \frac{268}{9} 
    - \frac{4\pi^2}{3} \right) C_A - \frac{40}{9}\, n_f \,,
    \nonumber\\
   \gamma_2^\text{cusp} &=& C_A^2 \left( \frac{490}{3} 
    - \frac{536\pi^2}{27}
    + \frac{44\pi^4}{45} + \frac{88}{3}\,\zeta_3 \right) 
    + C_A  n_f  \left( - \frac{836}{27} + \frac{80\pi^2}{27}
    - \frac{112}{3}\,\zeta_3 \right) \nonumber\\
   &&\mbox{}+ C_F n_f \left( - \frac{110}{3} + 32\zeta_3 \right) 
    - \frac{16}{27}\, n_f^2 \; ,
\end{eqnarray}
and for the quark anomalous dimension~\cite{Moch:2005id,Moch:2005tm}
\begin{eqnarray}
   \gamma_0^q &=& -3 C_F \,, \nonumber\\
   \gamma_1^q &=& C_F^2 \left( -\frac{3}{2} + 2\pi^2
    - 24\zeta_3 \right)
    + C_F C_A \left( - \frac{961}{54} - \frac{11\pi^2}{6} 
    + 26\zeta_3 \right)
    + C_F  n_f \left( \frac{65}{27} + \frac{\pi^2}{3} \right) ,
    \nonumber\\
   \gamma_2^q &=& C_F^3 \left( -\frac{29}{2} - 3\pi^2
    - \frac{8\pi^4}{5}
    - 68\zeta_3 + \frac{16\pi^2}{3}\,\zeta_3 + 240\zeta_5 \right) 
    \nonumber\\
   &&\mbox{}+ C_F^2 C_A \left( - \frac{151}{4} + \frac{205\pi^2}{9}
    + \frac{247\pi^4}{135} - \frac{844}{3}\,\zeta_3
    - \frac{8\pi^2}{3}\,\zeta_3 - 120\zeta_5 \right) \nonumber\\
   &&\mbox{}+ C_F C_A^2 \left( - \frac{139345}{2916} - \frac{7163\pi^2}{486}
    - \frac{83\pi^4}{90} + \frac{3526}{9}\,\zeta_3
    - \frac{44\pi^2}{9}\,\zeta_3 - 136\zeta_5 \right) \nonumber\\
   &&\mbox{}+ C_F^2  n_f \left( \frac{2953}{54} - \frac{13\pi^2}{9} 
    - \frac{14\pi^4}{27} + \frac{256}{9}\,\zeta_3 \right) 
    \nonumber\\
   &&\mbox{}+ C_F C_A n_f \left( - \frac{8659}{729}
    + \frac{1297\pi^2}{243} + \frac{11\pi^4}{45} 
    - \frac{964}{27}\,\zeta_3 \right) \nonumber\\
   &&\mbox{}+ C_F  n_f^2 \left( \frac{2417}{729} 
    - \frac{10\pi^2}{27} - \frac{8}{27}\,\zeta_3 \right) \; ,
\end{eqnarray}
where we have set $T_F = \frac{1}{2}$.

%% file: 11-appC.tex

\section{Analytic Continuation} \label{sec:appC}
\begin{figure}
\center
\includegraphics[scale=0.3]{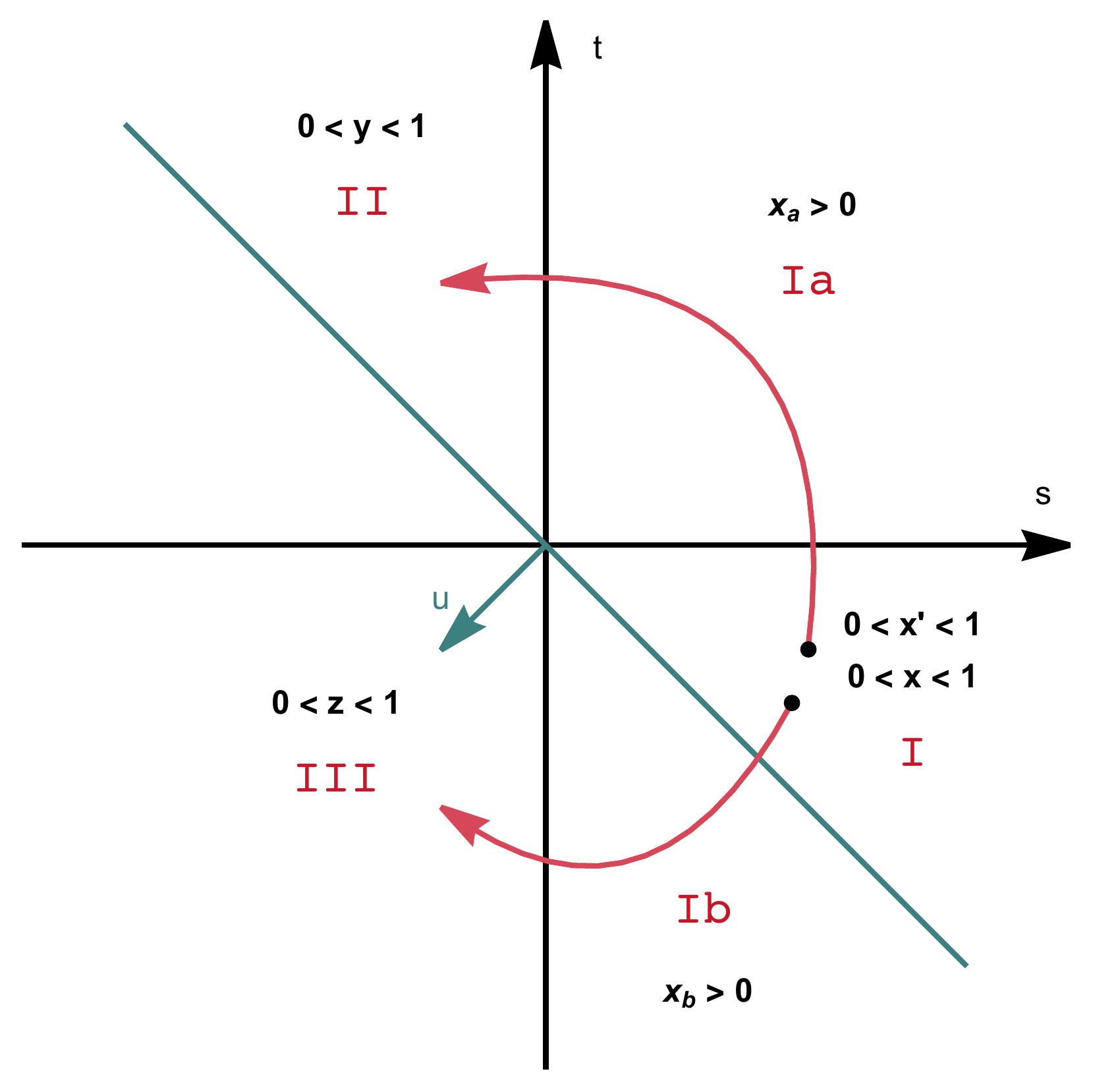}
\caption{Paths in phase space to obtain the crossed amplitudes.}
\label{KinematicsPath}
\end{figure}

In this appendix, we briefly outline the analytical continuation of the scattering amplitudes needed to obtain the crossed channels presented in section~\ref{extra_results}. The kinematic regions and the connecting paths used in the following are shown in figure~\ref{KinematicsPath}.
\paragraph{Process $ q Q \rightarrow  q Q$:} To derive the partonic channel~\eqref{t_channel} from our calculation of process~\eqref{s_channel_new}, one must effectively continue our results from region \texttt{I} to region \texttt{II}, where $p_1$ and $p_3$ are incoming, see figure \ref{KinematicsPath}. After the analytic continuation, one can then rename $p_2 \leftrightarrow p_3$.  The connecting path crosses two branch cuts, one at $t=0$ and one at $s=0$.  
After crossing each branch cut, we wish to maintain the amplitude written in terms of explicitly real HPLs.
This is achieved by the change of variable $x \rightarrow -x_a - i\delta$ for the first branch cut and $x_a \rightarrow -1/y -i \delta$ for the second, where the variables used have been defined in eqs.~\eqref{variables}.
We stress here, that the sign of the infinitesimal imaginary parts are chosen by imposing
that each Mandelstam invariant that changes sign effectively carries a small \emph{positive} imaginary part.
Once we are in region \texttt{II}, with the amplitude written as a function of $y$, 
we can perform the renaming  $p_2 \leftrightarrow p_3$,
\begin{equation}
y = x|_{p_2 \leftrightarrow p_3} \; \xrightarrow{p_2 \leftrightarrow p_3} \;x \; ,
\end{equation}
to match the usual conventions.
\paragraph{Process $ q \bar Q \rightarrow  q \bar Q$:} To obtain process \eqref{u_channel} from process \eqref{s_channel_new}, we move from region \texttt{I} to region \texttt{III},  and then exchange  $p_1$ with $p_4$.  This time, we first cross the branch cut at $u=0$, and then the one at $s=0$.  It is convenient to perform a preliminary change of variables $x =1-x'$ prior to crossing the cuts.  We can then use exactly the same changes of variables as for the previous case: to cross the $u=0$ branch cut we use $x' \rightarrow -x_b - i\delta$, while for the $s=0$ branch cut we use $x_b \rightarrow -1/z -i \delta$, such that $0<z<1$ in region \texttt{III}. We can then perform the renaming $p_2 \leftrightarrow p_4$,
\begin{equation}
z = x|_{p_2 \leftrightarrow p_4} \; \xrightarrow{p_2 \leftrightarrow p_4} \; x \; .
\end{equation}
As for the previous case, the signs of the imaginary parts are always chosen
such that the Mandelstam invariants that change sign 
effectively carry a small \emph{positive} imaginary part.
We have performed all analytic continuations explicitly and employed \texttt{PolyLogTools}~\cite{Duhr:2019tlz} for the required transformations of the harmonic polylogarithms.

%% file: 0-qqQQmain.bbl
\providecommand{\href}[2]{#2}\begingroup\raggedright\begin{thebibliography}{100}

\bibitem{Tkachov:1981wb}
F.~Tkachov, \emph{{A Theorem on Analytical Calculability of Four Loop
  Renormalization Group Functions}},
  \href{https://doi.org/10.1016/0370-2693(81)90288-4}{\emph{Phys.Lett.}
  {\bfseries B100} (1981) 65}.

\bibitem{Chetyrkin:1981qh}
K.~Chetyrkin and F.~Tkachov, \emph{{Integration by Parts: The Algorithm to
  Calculate beta Functions in 4 Loops}},
  \href{https://doi.org/10.1016/0550-3213(81)90199-1}{\emph{Nucl.Phys.}
  {\bfseries B192} (1981) 159}.

\bibitem{Hodges:2009hk}
A.~Hodges, \emph{{Eliminating spurious poles from gauge-theoretic amplitudes}},
  \href{https://doi.org/10.1007/JHEP05(2013)135}{\emph{JHEP} {\bfseries 05}
  (2013) 135} [\href{https://arxiv.org/abs/0905.1473}{{\ttfamily 0905.1473}}].

\bibitem{Gluza:2010ws}
J.~Gluza, K.~Kajda and D.~A. Kosower, \emph{{Towards a Basis for Planar
  Two-Loop Integrals}},
  \href{https://doi.org/10.1103/PhysRevD.83.045012}{\emph{Phys. Rev. D}
  {\bfseries 83} (2011) 045012}
  [\href{https://arxiv.org/abs/1009.0472}{{\ttfamily 1009.0472}}].

\bibitem{Ita:2015tya}
H.~Ita, \emph{{Two-loop Integrand Decomposition into Master Integrals and
  Surface Terms}},
  \href{https://doi.org/10.1103/PhysRevD.94.116015}{\emph{Phys. Rev.}
  {\bfseries D94} (2016) 116015}
  [\href{https://arxiv.org/abs/1510.05626}{{\ttfamily 1510.05626}}].

\bibitem{Larsen:2015ped}
K.~J. Larsen and Y.~Zhang, \emph{{Integration-by-parts reductions from
  unitarity cuts and algebraic geometry}},
  \href{https://doi.org/10.1103/PhysRevD.93.041701}{\emph{Phys. Rev.}
  {\bfseries D93} (2016) 041701}
  [\href{https://arxiv.org/abs/1511.01071}{{\ttfamily 1511.01071}}].

\bibitem{Bohm:2017qme}
J.~B\"ohm, A.~Georgoudis, K.~J. Larsen, M.~Schulze and Y.~Zhang,
  \emph{{Complete sets of logarithmic vector fields for integration-by-parts
  identities of Feynman integrals}},
  \href{https://doi.org/10.1103/PhysRevD.98.025023}{\emph{Phys. Rev. D}
  {\bfseries 98} (2018) 025023}
  [\href{https://arxiv.org/abs/1712.09737}{{\ttfamily 1712.09737}}].

\bibitem{Badger:2016uuq}
S.~Badger, \emph{{Automating QCD amplitudes with on-shell methods}},
  \href{https://doi.org/10.1088/1742-6596/762/1/012057}{\emph{J. Phys. Conf.
  Ser.} {\bfseries 762} (2016) 012057}
  [\href{https://arxiv.org/abs/1605.02172}{{\ttfamily 1605.02172}}].

\bibitem{vonManteuffel:2014ixa}
A.~von Manteuffel and R.~M. Schabinger, \emph{{A novel approach to integration
  by parts reduction}},
  \href{https://doi.org/10.1016/j.physletb.2015.03.029}{\emph{Phys. Lett.}
  {\bfseries B744} (2015) 101}
  [\href{https://arxiv.org/abs/1406.4513}{{\ttfamily 1406.4513}}].

\bibitem{Peraro:2016wsq}
T.~Peraro, \emph{{Scattering amplitudes over finite fields and multivariate
  functional reconstruction}},
  \href{https://doi.org/10.1007/JHEP12(2016)030}{\emph{JHEP} {\bfseries 12}
  (2016) 030} [\href{https://arxiv.org/abs/1608.01902}{{\ttfamily
  1608.01902}}].

\bibitem{Peraro:2019svx}
T.~Peraro, \emph{{FiniteFlow: multivariate functional reconstruction using
  finite fields and dataflow graphs}},
  \href{https://arxiv.org/abs/1905.08019}{{\ttfamily 1905.08019}}.

\bibitem{Guan:2019bcx}
X.~Guan, X.~Liu and Y.-Q. Ma, \emph{{Complete reduction of integrals in
  two-loop five-light-parton scattering amplitudes}},
  \href{https://doi.org/10.1088/1674-1137/44/9/093106}{\emph{Chin. Phys. C}
  {\bfseries 44} (2020) 093106}
  [\href{https://arxiv.org/abs/1912.09294}{{\ttfamily 1912.09294}}].

\bibitem{Pak:2011xt}
A.~Pak, \emph{{The Toolbox of modern multi-loop calculations: novel analytic
  and semi-analytic techniques}},
  \href{https://doi.org/10.1088/1742-6596/368/1/012049}{\emph{J. Phys. Conf.
  Ser.} {\bfseries 368} (2012) 012049}
  [\href{https://arxiv.org/abs/1111.0868}{{\ttfamily 1111.0868}}].

\bibitem{Abreu:2019odu}
S.~Abreu, J.~Dormans, F.~Febres~Cordero, H.~Ita, B.~Page and V.~Sotnikov,
  \emph{{Analytic Form of the Planar Two-Loop Five-Parton Scattering Amplitudes
  in QCD}}, \href{https://doi.org/10.1007/JHEP05(2019)084}{\emph{JHEP}
  {\bfseries 05} (2019) 084}
  [\href{https://arxiv.org/abs/1904.00945}{{\ttfamily 1904.00945}}].

\bibitem{Heller:2021qkz}
M.~Heller and A.~von Manteuffel, \emph{{MultivariateApart: Generalized Partial
  Fractions}},  \href{https://arxiv.org/abs/2101.08283}{{\ttfamily
  2101.08283}}.

\bibitem{Kotikov:1990kg}
A.~Kotikov, \emph{{Differential equations method: New technique for massive
  Feynman diagrams calculation}},
  \href{https://doi.org/10.1016/0370-2693(91)90413-K}{\emph{Phys.Lett.}
  {\bfseries B254} (1991) 158}.

\bibitem{Bern:1993kr}
Z.~Bern, L.~J. Dixon and D.~A. Kosower, \emph{{Dimensionally regulated pentagon
  integrals}}, \href{https://doi.org/10.1016/0550-3213(94)90398-0}{\emph{Nucl.
  Phys.} {\bfseries B412} (1994) 751}
  [\href{https://arxiv.org/abs/hep-ph/9306240}{{\ttfamily hep-ph/9306240}}].

\bibitem{Remiddi:1997ny}
E.~Remiddi, \emph{{Differential equations for Feynman graph amplitudes}},
  {\emph{Nuovo Cim.} {\bfseries A110} (1997) 1435}
  [\href{https://arxiv.org/abs/hep-th/9711188}{{\ttfamily hep-th/9711188}}].

\bibitem{Gehrmann:1999as}
T.~Gehrmann and E.~Remiddi, \emph{{Differential equations for two loop four
  point functions}},
  \href{https://doi.org/10.1016/S0550-3213(00)00223-6}{\emph{Nucl.Phys.}
  {\bfseries B580} (2000) 485}
  [\href{https://arxiv.org/abs/hep-ph/9912329}{{\ttfamily hep-ph/9912329}}].

\bibitem{Papadopoulos:2014lla}
C.~G. Papadopoulos, \emph{{Simplified differential equations approach for
  Master Integrals}},
  \href{https://doi.org/10.1007/JHEP07(2014)088}{\emph{JHEP} {\bfseries 07}
  (2014) 088} [\href{https://arxiv.org/abs/1401.6057}{{\ttfamily 1401.6057}}].

\bibitem{Dixon:1996wi}
L.~J. Dixon, \emph{{Calculating scattering amplitudes efficiently}},
  \href{https://arxiv.org/abs/hep-ph/9601359}{{\ttfamily hep-ph/9601359}}.

\bibitem{Henn:2013pwa}
J.~M. Henn, \emph{{Multiloop integrals in dimensional regularization made
  simple}},
  \href{https://doi.org/10.1103/PhysRevLett.110.251601}{\emph{Phys.Rev.Lett.}
  {\bfseries 110} (2013) 251601}
  [\href{https://arxiv.org/abs/1304.1806}{{\ttfamily 1304.1806}}].

\bibitem{Primo:2016ebd}
A.~Primo and L.~Tancredi, \emph{{On the maximal cut of Feynman integrals and
  the solution of their differential equations}},
  \href{https://doi.org/10.1016/j.nuclphysb.2016.12.021}{\emph{Nucl. Phys.}
  {\bfseries B916} (2017) 94}
  [\href{https://arxiv.org/abs/1610.08397}{{\ttfamily 1610.08397}}].

\bibitem{Goncharov}
A.~B. Goncharov, \emph{{Geometry of configurations, polylogarithms, and motivic
  cohomology}}, \href{https://doi.org/10.1006/aima.1995.1045}{\emph{Adv. Math.}
  {\bfseries 114} (1995) 197}.

\bibitem{Remiddi:1999ew}
E.~Remiddi and J.~Vermaseren, \emph{{Harmonic polylogarithms}},
  \href{https://doi.org/10.1142/S0217751X00000367}{\emph{Int.J.Mod.Phys.}
  {\bfseries A15} (2000) 725}
  [\href{https://arxiv.org/abs/hep-ph/9905237}{{\ttfamily hep-ph/9905237}}].

\bibitem{Goncharov:2001iea}
A.~B. Goncharov, \emph{{Multiple polylogarithms and mixed Tate motives}},
  \href{https://arxiv.org/abs/math/0103059}{{\ttfamily math/0103059}}.

\bibitem{Goncharov:2010jf}
A.~B. Goncharov, M.~Spradlin, C.~Vergu and A.~Volovich, \emph{{Classical
  Polylogarithms for Amplitudes and Wilson Loops}},
  \href{https://doi.org/10.1103/PhysRevLett.105.151605}{\emph{Phys.Rev.Lett.}
  {\bfseries 105} (2010) 151605}
  [\href{https://arxiv.org/abs/1006.5703}{{\ttfamily 1006.5703}}].

\bibitem{Brown:2008um}
F.~Brown, \emph{{The Massless higher-loop two-point function}},
  \href{https://doi.org/10.1007/s00220-009-0740-5}{\emph{Commun.Math.Phys.}
  {\bfseries 287} (2009) 925}
  [\href{https://arxiv.org/abs/0804.1660}{{\ttfamily 0804.1660}}].

\bibitem{Ablinger:2013cf}
J.~Ablinger, J.~Bl{\"u}mlein and C.~Schneider, \emph{{Analytic and Algorithmic
  Aspects of Generalized Harmonic Sums and Polylogarithms}},
  \href{https://doi.org/10.1063/1.4811117}{\emph{J. Math. Phys.} {\bfseries 54}
  (2013) 082301} [\href{https://arxiv.org/abs/1302.0378}{{\ttfamily
  1302.0378}}].

\bibitem{Panzer:2014caa}
E.~Panzer, \emph{{Algorithms for the symbolic integration of hyperlogarithms
  with applications to Feynman integrals}},
  \href{https://doi.org/10.1016/j.cpc.2014.10.019}{\emph{Comput.Phys.Commun.}
  {\bfseries 188} (2014) 148}
  [\href{https://arxiv.org/abs/1403.3385}{{\ttfamily 1403.3385}}].

\bibitem{Duhr:2011zq}
C.~Duhr, H.~Gangl and J.~R. Rhodes, \emph{{From polygons and symbols to
  polylogarithmic functions}},
  \href{https://doi.org/10.1007/JHEP10(2012)075}{\emph{JHEP} {\bfseries 1210}
  (2012) 075} [\href{https://arxiv.org/abs/1110.0458}{{\ttfamily 1110.0458}}].

\bibitem{Duhr:2012fh}
C.~Duhr, \emph{{Hopf algebras, coproducts and symbols: an application to Higgs
  boson amplitudes}},
  \href{https://doi.org/10.1007/JHEP08(2012)043}{\emph{JHEP} {\bfseries 1208}
  (2012) 043} [\href{https://arxiv.org/abs/1203.0454}{{\ttfamily 1203.0454}}].

\bibitem{Duhr:2019tlz}
C.~Duhr and F.~Dulat, \emph{{PolyLogTools \textemdash{} polylogs for the
  masses}}, \href{https://doi.org/10.1007/JHEP08(2019)135}{\emph{JHEP}
  {\bfseries 08} (2019) 135}
  [\href{https://arxiv.org/abs/1904.07279}{{\ttfamily 1904.07279}}].

\bibitem{Badger:2017jhb}
S.~Badger, C.~Broennum-Hansen, H.~B. Hartanto and T.~Peraro, \emph{{First look
  at two-loop five-gluon scattering in QCD}},
  \href{https://doi.org/10.1103/PhysRevLett.120.092001}{\emph{Phys. Rev. Lett.}
  {\bfseries 120} (2018) 092001}
  [\href{https://arxiv.org/abs/1712.02229}{{\ttfamily 1712.02229}}].

\bibitem{Abreu:2017hqn}
S.~Abreu, F.~Febres~Cordero, H.~Ita, B.~Page and M.~Zeng, \emph{{Planar
  Two-Loop Five-Gluon Amplitudes from Numerical Unitarity}},
  \href{https://doi.org/10.1103/PhysRevD.97.116014}{\emph{Phys. Rev.}
  {\bfseries D97} (2018) 116014}
  [\href{https://arxiv.org/abs/1712.03946}{{\ttfamily 1712.03946}}].

\bibitem{Abreu:2018aqd}
S.~Abreu, L.~J. Dixon, E.~Herrmann, B.~Page and M.~Zeng, \emph{{The two-loop
  five-point amplitude in $\mathcal{N} =4$ super-Yang-Mills theory}},
  \href{https://doi.org/10.1103/PhysRevLett.122.121603}{\emph{Phys. Rev. Lett.}
  {\bfseries 122} (2019) 121603}
  [\href{https://arxiv.org/abs/1812.08941}{{\ttfamily 1812.08941}}].

\bibitem{Abreu:2018zmy}
S.~Abreu, J.~Dormans, F.~Febres~Cordero, H.~Ita and B.~Page, \emph{{Analytic
  Form of Planar Two-Loop Five-Gluon Scattering Amplitudes in QCD}},
  \href{https://doi.org/10.1103/PhysRevLett.122.082002}{\emph{Phys. Rev. Lett.}
  {\bfseries 122} (2019) 082002}
  [\href{https://arxiv.org/abs/1812.04586}{{\ttfamily 1812.04586}}].

\bibitem{Abreu:2018jgq}
S.~Abreu, F.~Febres~Cordero, H.~Ita, B.~Page and V.~Sotnikov, \emph{{Planar
  Two-Loop Five-Parton Amplitudes from Numerical Unitarity}},
  \href{https://doi.org/10.1007/JHEP11(2018)116}{\emph{JHEP} {\bfseries 11}
  (2018) 116} [\href{https://arxiv.org/abs/1809.09067}{{\ttfamily
  1809.09067}}].

\bibitem{Abreu:2019rpt}
S.~Abreu, L.~J. Dixon, E.~Herrmann, B.~Page and M.~Zeng, \emph{{The two-loop
  five-point amplitude in $ \mathcal{N} $ = 8 supergravity}},
  \href{https://doi.org/10.1007/JHEP03(2019)123}{\emph{JHEP} {\bfseries 03}
  (2019) 123} [\href{https://arxiv.org/abs/1901.08563}{{\ttfamily
  1901.08563}}].

\bibitem{Abreu:2020cwb}
S.~Abreu, B.~Page, E.~Pascual and V.~Sotnikov, \emph{{Leading-Color Two-Loop
  QCD Corrections for Three-Photon Production at Hadron Colliders}},
  \href{https://doi.org/10.1007/JHEP01(2021)078}{\emph{JHEP} {\bfseries 01}
  (2021) 078} [\href{https://arxiv.org/abs/2010.15834}{{\ttfamily
  2010.15834}}].

\bibitem{Chicherin:2018yne}
D.~Chicherin, T.~Gehrmann, J.~M. Henn, P.~Wasser, Y.~Zhang and S.~Zoia,
  \emph{{Analytic result for a two-loop five-particle amplitude}},
  \href{https://doi.org/10.1103/PhysRevLett.122.121602}{\emph{Phys. Rev. Lett.}
  {\bfseries 122} (2019) 121602}
  [\href{https://arxiv.org/abs/1812.11057}{{\ttfamily 1812.11057}}].

\bibitem{Chicherin:2019xeg}
D.~Chicherin, T.~Gehrmann, J.~M. Henn, P.~Wasser, Y.~Zhang and S.~Zoia,
  \emph{{The two-loop five-particle amplitude in $ \mathcal{N} $ = 8
  supergravity}}, \href{https://doi.org/10.1007/JHEP03(2019)115}{\emph{JHEP}
  {\bfseries 03} (2019) 115}
  [\href{https://arxiv.org/abs/1901.05932}{{\ttfamily 1901.05932}}].

\bibitem{Chawdhry:2020for}
H.~A. Chawdhry, M.~Czakon, A.~Mitov and R.~Poncelet, \emph{{Two-loop
  leading-color helicity amplitudes for three-photon production at the LHC}},
  \href{https://arxiv.org/abs/2012.13553}{{\ttfamily 2012.13553}}.

\bibitem{DeLaurentis:2020qle}
G.~De~Laurentis and D.~Ma\^\i{}tre, \emph{{Two-Loop Five-Parton Leading-Colour
  Finite Remainders in the Spinor-Helicity Formalism}},
  \href{https://doi.org/10.1007/JHEP02(2021)016}{\emph{JHEP} {\bfseries 02}
  (2021) 016} [\href{https://arxiv.org/abs/2010.14525}{{\ttfamily
  2010.14525}}].

\bibitem{Chawdhry:2018awn}
H.~A. Chawdhry, M.~A. Lim and A.~Mitov, \emph{{Two-loop five-point massless QCD
  amplitudes within the integration-by-parts approach}},
  \href{https://doi.org/10.1103/PhysRevD.99.076011}{\emph{Phys. Rev. D}
  {\bfseries 99} (2019) 076011}
  [\href{https://arxiv.org/abs/1805.09182}{{\ttfamily 1805.09182}}].

\bibitem{Abreu:2020xvt}
S.~Abreu, J.~Dormans, F.~Febres~Cordero, H.~Ita, M.~Kraus, B.~Page et~al.,
  \emph{{Caravel: A C++ framework for the computation of multi-loop amplitudes
  with numerical unitarity}},
  \href{https://doi.org/10.1016/j.cpc.2021.108069}{\emph{Comput. Phys. Commun.}
  {\bfseries 267} (2021) 108069}
  [\href{https://arxiv.org/abs/2009.11957}{{\ttfamily 2009.11957}}].

\bibitem{Agarwal:2021grm}
B.~Agarwal, F.~Buccioni, A.~von Manteuffel and L.~Tancredi, \emph{{Two-loop
  leading colour QCD corrections to $q \bar{q} \to \gamma \gamma g$ and $q g
  \to \gamma \gamma q$}},
  \href{https://doi.org/10.1007/JHEP04(2021)201}{\emph{JHEP} {\bfseries 04}
  (2021) 201} [\href{https://arxiv.org/abs/2102.01820}{{\ttfamily
  2102.01820}}].

\bibitem{Badger:2021nhg}
S.~Badger, H.~B. Hartanto and S.~Zoia, \emph{{Two-loop QCD corrections to
  $Wb\bar{b}$ production at hadron colliders}},
  \href{https://arxiv.org/abs/2102.02516}{{\ttfamily 2102.02516}}.

\bibitem{Abreu:2021fuk}
S.~Abreu, F.~F. Cordero, H.~Ita, B.~Page and V.~Sotnikov, \emph{{Leading-Color
  Two-Loop QCD Corrections for Three-Jet Production at Hadron Colliders}},
  \href{https://arxiv.org/abs/2102.13609}{{\ttfamily 2102.13609}}.

\bibitem{Agarwal:2021vdh}
B.~Agarwal, F.~Buccioni, A.~von Manteuffel and L.~Tancredi, \emph{{Two-loop
  helicity amplitudes for diphoton plus jet production in full color}},
  \href{https://arxiv.org/abs/2105.04585}{{\ttfamily 2105.04585}}.

\bibitem{Chawdhry:2021mkw}
H.~A. Chawdhry, M.~Czakon, A.~Mitov and R.~Poncelet, \emph{{Two-loop
  leading-colour QCD helicity amplitudes for two-photon plus jet production at
  the LHC}},  \href{https://arxiv.org/abs/2103.04319}{{\ttfamily 2103.04319}}.

\bibitem{Badger:2021imn}
S.~Badger, C.~Br\o{}nnum-Hansen, D.~Chicherin, T.~Gehrmann, H.~B. Hartanto,
  J.~Henn et~al., \emph{{Virtual QCD corrections to gluon-initiated diphoton
  plus jet production at hadron colliders}},
  \href{https://arxiv.org/abs/2106.08664}{{\ttfamily 2106.08664}}.

\bibitem{Gehrmann:2015bfy}
T.~Gehrmann, J.~M. Henn and N.~A. Lo~Presti, \emph{{Analytic form of the
  two-loop planar five-gluon all-plus-helicity amplitude in QCD}},
  \href{https://doi.org/10.1103/PhysRevLett.116.062001}{\emph{Phys. Rev. Lett.}
  {\bfseries 116} (2016) 062001}
  [\href{https://arxiv.org/abs/1511.05409}{{\ttfamily 1511.05409}}].

\bibitem{Papadopoulos:2015jft}
C.~G. Papadopoulos, D.~Tommasini and C.~Wever, \emph{{The Pentabox Master
  Integrals with the Simplified Differential Equations approach}},
  \href{https://doi.org/10.1007/JHEP04(2016)078}{\emph{JHEP} {\bfseries 04}
  (2016) 078} [\href{https://arxiv.org/abs/1511.09404}{{\ttfamily
  1511.09404}}].

\bibitem{Gehrmann:2018yef}
T.~Gehrmann, J.~Henn and N.~Lo~Presti, \emph{{Pentagon functions for massless
  planar scattering amplitudes}},
  \href{https://doi.org/10.1007/JHEP10(2018)103}{\emph{JHEP} {\bfseries 10}
  (2018) 103} [\href{https://arxiv.org/abs/1807.09812}{{\ttfamily
  1807.09812}}].

\bibitem{Chicherin:2018mue}
D.~Chicherin, T.~Gehrmann, J.~Henn, N.~Lo~Presti, V.~Mitev and P.~Wasser,
  \emph{{Analytic result for the nonplanar hexa-box integrals}},
  \href{https://doi.org/10.1007/JHEP03(2019)042}{\emph{JHEP} {\bfseries 03}
  (2019) 042} [\href{https://arxiv.org/abs/1809.06240}{{\ttfamily
  1809.06240}}].

\bibitem{Chicherin:2020oor}
D.~Chicherin and V.~Sotnikov, \emph{{Pentagon Functions for Scattering of Five
  Massless Particles}},
  \href{https://doi.org/10.1007/JHEP12(2020)167}{\emph{JHEP} {\bfseries 20}
  (2020) 167} [\href{https://arxiv.org/abs/2009.07803}{{\ttfamily
  2009.07803}}].

\bibitem{Chawdhry:2019bji}
H.~A. Chawdhry, M.~L. Czakon, A.~Mitov and R.~Poncelet, \emph{{NNLO QCD
  corrections to three-photon production at the LHC}},
  \href{https://doi.org/10.1007/JHEP02(2020)057}{\emph{JHEP} {\bfseries 02}
  (2020) 057} [\href{https://arxiv.org/abs/1911.00479}{{\ttfamily
  1911.00479}}].

\bibitem{Kallweit:2020gcp}
S.~Kallweit, V.~Sotnikov and M.~Wiesemann, \emph{{Triphoton production at
  hadron colliders in NNLO QCD}},
  \href{https://doi.org/10.1016/j.physletb.2020.136013}{\emph{Phys. Lett. B}
  {\bfseries 812} (2021) 136013}
  [\href{https://arxiv.org/abs/2010.04681}{{\ttfamily 2010.04681}}].

\bibitem{Chawdhry:2021hkp}
H.~A. Chawdhry, M.~Czakon, A.~Mitov and R.~Poncelet, \emph{{NNLO QCD
  corrections to diphoton production with an additional jet at the LHC}},
  \href{https://arxiv.org/abs/2105.06940}{{\ttfamily 2105.06940}}.

\bibitem{Czakon:2021mjy}
M.~Czakon, A.~Mitov and R.~Poncelet, \emph{{Tour de force in Quantum
  Chromodynamics: A first next-to-next-to-leading order study of three-jet
  production at the LHC}},  \href{https://arxiv.org/abs/2106.05331}{{\ttfamily
  2106.05331}}.

\bibitem{GehrmannDeRidder:2005cm}
A.~Gehrmann-De~Ridder, T.~Gehrmann and E.~N. Glover, \emph{{Antenna subtraction
  at NNLO}}, \href{https://doi.org/10.1088/1126-6708/2005/09/056}{\emph{JHEP}
  {\bfseries 0509} (2005) 056}
  [\href{https://arxiv.org/abs/hep-ph/0505111}{{\ttfamily hep-ph/0505111}}].

\bibitem{Czakon:2010td}
M.~Czakon, \emph{{A novel subtraction scheme for double-real radiation at
  NNLO}}, \href{https://doi.org/10.1016/j.physletb.2010.08.036}{\emph{Phys.
  Lett. B} {\bfseries 693} (2010) 259}
  [\href{https://arxiv.org/abs/1005.0274}{{\ttfamily 1005.0274}}].

\bibitem{Caola:2017dug}
F.~Caola, K.~Melnikov and R.~R\"ontsch, \emph{{Nested soft-collinear
  subtractions in NNLO QCD computations}},
  \href{https://doi.org/10.1140/epjc/s10052-017-4774-0}{\emph{Eur. Phys. J. C}
  {\bfseries 77} (2017) 248}
  [\href{https://arxiv.org/abs/1702.01352}{{\ttfamily 1702.01352}}].

\bibitem{Magnea:2018hab}
L.~Magnea, E.~Maina, G.~Pelliccioli, C.~Signorile-Signorile, P.~Torrielli and
  S.~Uccirati, \emph{{Local analytic sector subtraction at NNLO}},
  \href{https://doi.org/10.1007/JHEP12(2018)107}{\emph{JHEP} {\bfseries 12}
  (2018) 107} [\href{https://arxiv.org/abs/1806.09570}{{\ttfamily
  1806.09570}}].

\bibitem{Herzog:2018ily}
F.~Herzog, \emph{{Geometric IR subtraction for final state real radiation}},
  \href{https://doi.org/10.1007/JHEP08(2018)006}{\emph{JHEP} {\bfseries 08}
  (2018) 006} [\href{https://arxiv.org/abs/1804.07949}{{\ttfamily
  1804.07949}}].

\bibitem{DelDuca:2016ily}
V.~Del~Duca, C.~Duhr, A.~Kardos, G.~Somogyi, Z.~Sz\H{o}r, Z.~Tr\'ocs\'anyi
  et~al., \emph{{Jet production in the CoLoRFulNNLO method: event shapes in
  electron-positron collisions}},
  \href{https://doi.org/10.1103/PhysRevD.94.074019}{\emph{Phys. Rev. D}
  {\bfseries 94} (2016) 074019}
  [\href{https://arxiv.org/abs/1606.03453}{{\ttfamily 1606.03453}}].

\bibitem{Cacciari:2015jma}
M.~Cacciari, F.~A. Dreyer, A.~Karlberg, G.~P. Salam and G.~Zanderighi,
  \emph{{Fully Differential Vector-Boson-Fusion Higgs Production at
  Next-to-Next-to-Leading Order}},
  \href{https://doi.org/10.1103/PhysRevLett.115.082002}{\emph{Phys. Rev. Lett.}
  {\bfseries 115} (2015) 082002}
  [\href{https://arxiv.org/abs/1506.02660}{{\ttfamily 1506.02660}}].

\bibitem{Catani:2007vq}
S.~Catani and M.~Grazzini, \emph{{An NNLO subtraction formalism in hadron
  collisions and its application to Higgs boson production at the LHC}},
  \href{https://doi.org/10.1103/PhysRevLett.98.222002}{\emph{Phys.Rev.Lett.}
  {\bfseries 98} (2007) 222002}
  [\href{https://arxiv.org/abs/hep-ph/0703012}{{\ttfamily hep-ph/0703012}}].

\bibitem{Gaunt:2015pea}
J.~Gaunt, M.~Stahlhofen, F.~J. Tackmann and J.~R. Walsh, \emph{{N-jettiness
  Subtractions for NNLO QCD Calculations}},
  \href{https://doi.org/10.1007/JHEP09(2015)058}{\emph{JHEP} {\bfseries 09}
  (2015) 058} [\href{https://arxiv.org/abs/1505.04794}{{\ttfamily
  1505.04794}}].

\bibitem{Boughezal:2015dva}
R.~Boughezal, C.~Focke, X.~Liu and F.~Petriello, \emph{{$W$-boson production in
  association with a jet at next-to-next-to-leading order in perturbative
  QCD}}, \href{https://doi.org/10.1103/PhysRevLett.115.062002}{\emph{Phys. Rev.
  Lett.} {\bfseries 115} (2015) 062002}
  [\href{https://arxiv.org/abs/1504.02131}{{\ttfamily 1504.02131}}].

\bibitem{Anastasiou:2015vya}
C.~Anastasiou, C.~Duhr, F.~Dulat, F.~Herzog and B.~Mistlberger, \emph{{Higgs
  Boson Gluon-Fusion Production in QCD at Three Loops}},
  \href{https://doi.org/10.1103/PhysRevLett.114.212001}{\emph{Phys. Rev. Lett.}
  {\bfseries 114} (2015) 212001}
  [\href{https://arxiv.org/abs/1503.06056}{{\ttfamily 1503.06056}}].

\bibitem{Duhr:2019kwi}
C.~Duhr, F.~Dulat and B.~Mistlberger, \emph{{Higgs Boson Production in
  Bottom-Quark Fusion to Third Order in the Strong Coupling}},
  \href{https://doi.org/10.1103/PhysRevLett.125.051804}{\emph{Phys. Rev. Lett.}
  {\bfseries 125} (2020) 051804}
  [\href{https://arxiv.org/abs/1904.09990}{{\ttfamily 1904.09990}}].

\bibitem{Dulat:2018bfe}
F.~Dulat, B.~Mistlberger and A.~Pelloni, \emph{{Precision predictions at
  N$^3$LO for the Higgs boson rapidity distribution at the LHC}},
  \href{https://doi.org/10.1103/PhysRevD.99.034004}{\emph{Phys. Rev. D}
  {\bfseries 99} (2019) 034004}
  [\href{https://arxiv.org/abs/1810.09462}{{\ttfamily 1810.09462}}].

\bibitem{Mistlberger:2018etf}
B.~Mistlberger, \emph{{Higgs boson production at hadron colliders at N$^{3}$LO
  in QCD}}, \href{https://doi.org/10.1007/JHEP05(2018)028}{\emph{JHEP}
  {\bfseries 05} (2018) 028}
  [\href{https://arxiv.org/abs/1802.00833}{{\ttfamily 1802.00833}}].

\bibitem{Dreyer:2016oyx}
F.~A. Dreyer and A.~Karlberg, \emph{{Vector-Boson Fusion Higgs Production at
  Three Loops in QCD}},
  \href{https://doi.org/10.1103/PhysRevLett.117.072001}{\emph{Phys. Rev. Lett.}
  {\bfseries 117} (2016) 072001}
  [\href{https://arxiv.org/abs/1606.00840}{{\ttfamily 1606.00840}}].

\bibitem{Dreyer:2018qbw}
F.~A. Dreyer and A.~Karlberg, \emph{{Vector-Boson Fusion Higgs Pair Production
  at N$^3$LO}}, \href{https://doi.org/10.1103/PhysRevD.98.114016}{\emph{Phys.
  Rev. D} {\bfseries 98} (2018) 114016}
  [\href{https://arxiv.org/abs/1811.07906}{{\ttfamily 1811.07906}}].

\bibitem{Billis:2021ecs}
G.~Billis, B.~Dehnadi, M.~A. Ebert, J.~K.~L. Michel and F.~J. Tackmann,
  \emph{{The Higgs $p_T$ Spectrum and Total Cross Section with Fiducial Cuts at
  N$^3$LL$'$+N$^3$LO}},  \href{https://arxiv.org/abs/2102.08039}{{\ttfamily
  2102.08039}}.

\bibitem{Chen:2021isd}
X.~Chen, X.~Chen, T.~Gehrmann, E.~W.~N. Glover, A.~Huss, B.~Mistlberger et~al.,
  \emph{{Fully Differential Higgs Boson Production to Third Order in QCD}},
  \href{https://arxiv.org/abs/2102.07607}{{\ttfamily 2102.07607}}.

\bibitem{Chen:2021vtu}
X.~Chen, T.~Gehrmann, N.~Glover, A.~Huss, T.-Z. Yang and H.~X. Zhu,
  \emph{{Di-lepton Rapidity Distribution in Drell-Yan Production to Third Order
  in QCD}},  \href{https://arxiv.org/abs/2107.09085}{{\ttfamily 2107.09085}}.

\bibitem{Almelid:2015jia}
O.~Almelid, C.~Duhr and E.~Gardi, \emph{{Three-loop corrections to the soft
  anomalous dimension in multileg scattering}},
  \href{https://doi.org/10.1103/PhysRevLett.117.172002}{\emph{Phys. Rev. Lett.}
  {\bfseries 117} (2016) 172002}
  [\href{https://arxiv.org/abs/1507.00047}{{\ttfamily 1507.00047}}].

\bibitem{Sterman:2002qn}
G.~F. Sterman and M.~E. Tejeda-Yeomans, \emph{{Multiloop amplitudes and
  resummation}},
  \href{https://doi.org/10.1016/S0370-2693(02)03100-3}{\emph{Phys. Lett. B}
  {\bfseries 552} (2003) 48}
  [\href{https://arxiv.org/abs/hep-ph/0210130}{{\ttfamily hep-ph/0210130}}].

\bibitem{Aybat:2006wq}
S.~Mert~Aybat, L.~J. Dixon and G.~F. Sterman, \emph{{The Two-loop anomalous
  dimension matrix for soft gluon exchange}},
  \href{https://doi.org/10.1103/PhysRevLett.97.072001}{\emph{Phys. Rev. Lett.}
  {\bfseries 97} (2006) 072001}
  [\href{https://arxiv.org/abs/hep-ph/0606254}{{\ttfamily hep-ph/0606254}}].

\bibitem{Aybat:2006mz}
S.~Mert~Aybat, L.~J. Dixon and G.~F. Sterman, \emph{{The Two-loop soft
  anomalous dimension matrix and resummation at next-to-next-to leading pole}},
  \href{https://doi.org/10.1103/PhysRevD.74.074004}{\emph{Phys. Rev. D}
  {\bfseries 74} (2006) 074004}
  [\href{https://arxiv.org/abs/hep-ph/0607309}{{\ttfamily hep-ph/0607309}}].

\bibitem{Becher:2009cu}
T.~Becher and M.~Neubert, \emph{{Infrared singularities of scattering
  amplitudes in perturbative QCD}},
  \href{https://doi.org/10.1103/PhysRevLett.102.162001}{\emph{Phys. Rev. Lett.}
  {\bfseries 102} (2009) 162001}
  [\href{https://arxiv.org/abs/0901.0722}{{\ttfamily 0901.0722}}].

\bibitem{Gardi:2009qi}
E.~Gardi and L.~Magnea, \emph{{Factorization constraints for soft anomalous
  dimensions in QCD scattering amplitudes}},
  \href{https://doi.org/10.1088/1126-6708/2009/03/079}{\emph{JHEP} {\bfseries
  0903} (2009) 079} [\href{https://arxiv.org/abs/0901.1091}{{\ttfamily
  0901.1091}}].

\bibitem{Becher:2009qa}
T.~Becher and M.~Neubert, \emph{{On the Structure of Infrared Singularities of
  Gauge-Theory Amplitudes}},
  \href{https://doi.org/10.1088/1126-6708/2009/06/081}{\emph{JHEP} {\bfseries
  06} (2009) 081} [\href{https://arxiv.org/abs/0903.1126}{{\ttfamily
  0903.1126}}].

\bibitem{Dixon:2009gx}
L.~J. Dixon, \emph{{Matter Dependence of the Three-Loop Soft Anomalous
  Dimension Matrix}},
  \href{https://doi.org/10.1103/PhysRevD.79.091501}{\emph{Phys. Rev. D}
  {\bfseries 79} (2009) 091501}
  [\href{https://arxiv.org/abs/0901.3414}{{\ttfamily 0901.3414}}].

\bibitem{Catani:2011st}
S.~Catani, D.~de~Florian and G.~Rodrigo, \emph{{Space-like (versus time-like)
  collinear limits in QCD: Is factorization violated?}},
  \href{https://doi.org/10.1007/JHEP07(2012)026}{\emph{JHEP} {\bfseries 07}
  (2012) 026} [\href{https://arxiv.org/abs/1112.4405}{{\ttfamily 1112.4405}}].

\bibitem{Forshaw:2012bi}
J.~R. Forshaw, M.~H. Seymour and A.~Siodmok, \emph{{On the Breaking of
  Collinear Factorization in QCD}},
  \href{https://doi.org/10.1007/JHEP11(2012)066}{\emph{JHEP} {\bfseries 11}
  (2012) 066} [\href{https://arxiv.org/abs/1206.6363}{{\ttfamily 1206.6363}}].

\bibitem{Forshaw:2006fk}
J.~R. Forshaw, A.~Kyrieleis and M.~H. Seymour, \emph{{Super-leading logarithms
  in non-global observables in QCD}},
  \href{https://doi.org/10.1088/1126-6708/2006/08/059}{\emph{JHEP} {\bfseries
  08} (2006) 059} [\href{https://arxiv.org/abs/hep-ph/0604094}{{\ttfamily
  hep-ph/0604094}}].

\bibitem{Becher:2021zkk}
T.~Becher, M.~Neubert and D.~Y. Shao, \emph{{Resummation of Super-Leading
  Logarithms}},  \href{https://arxiv.org/abs/2107.01212}{{\ttfamily
  2107.01212}}.

\bibitem{Smirnov:1999gc}
V.~A. Smirnov, \emph{{Analytical result for dimensionally regularized massless
  on shell double box}},
  \href{https://doi.org/10.1016/S0370-2693(99)00777-7}{\emph{Phys. Lett. B}
  {\bfseries 460} (1999) 397}
  [\href{https://arxiv.org/abs/hep-ph/9905323}{{\ttfamily hep-ph/9905323}}].

\bibitem{Tausk:1999vh}
J.~Tausk, \emph{{Nonplanar massless two loop Feynman diagrams with four
  on-shell legs}},
  \href{https://doi.org/10.1016/S0370-2693(99)01277-0}{\emph{Phys.Lett.}
  {\bfseries B469} (1999) 225}
  [\href{https://arxiv.org/abs/hep-ph/9909506}{{\ttfamily hep-ph/9909506}}].

\bibitem{Glover:2001af}
E.~W.~N. Glover, C.~Oleari and M.~E. Tejeda-Yeomans, \emph{{Two loop QCD
  corrections to gluon-gluon scattering}},
  \href{https://doi.org/10.1016/S0550-3213(01)00210-3}{\emph{Nucl. Phys. B}
  {\bfseries 605} (2001) 467}
  [\href{https://arxiv.org/abs/hep-ph/0102201}{{\ttfamily hep-ph/0102201}}].

\bibitem{Anastasiou:2002zn}
C.~Anastasiou, E.~W.~N. Glover and M.~Tejeda-Yeomans, \emph{{Two loop QED and
  QCD corrections to massless fermion boson scattering}},
  \href{https://doi.org/10.1016/S0550-3213(02)00140-2}{\emph{Nucl.Phys.}
  {\bfseries B629} (2002) 255}
  [\href{https://arxiv.org/abs/hep-ph/0201274}{{\ttfamily hep-ph/0201274}}].

\bibitem{Glover:2003cm}
E.~Glover and M.~Tejeda-Yeomans, \emph{{Two loop QCD helicity amplitudes for
  massless quark massless gauge boson scattering}},
  \href{https://doi.org/10.1088/1126-6708/2003/06/033}{\emph{JHEP} {\bfseries
  06} (2003) 033} [\href{https://arxiv.org/abs/hep-ph/0304169}{{\ttfamily
  hep-ph/0304169}}].

\bibitem{Anastasiou:2000kg}
C.~Anastasiou, E.~W.~N. Glover, C.~Oleari and M.~E. Tejeda-Yeomans,
  \emph{{Two-loop QCD corrections to the scattering of massless distinct
  quarks}}, \href{https://doi.org/10.1016/S0550-3213(01)00079-7}{\emph{Nucl.
  Phys. B} {\bfseries 601} (2001) 318}
  [\href{https://arxiv.org/abs/hep-ph/0010212}{{\ttfamily hep-ph/0010212}}].

\bibitem{Bern:2003ck}
Z.~Bern, A.~De~Freitas and L.~J. Dixon, \emph{{Two loop helicity amplitudes for
  quark gluon scattering in QCD and gluino gluon scattering in supersymmetric
  Yang-Mills theory}},
  \href{https://doi.org/10.1088/1126-6708/2003/06/028}{\emph{JHEP} {\bfseries
  06} (2003) 028} [\href{https://arxiv.org/abs/hep-ph/0304168}{{\ttfamily
  hep-ph/0304168}}].

\bibitem{Glover:2004si}
E.~Glover, \emph{{Two loop QCD helicity amplitudes for massless quark quark
  scattering}},
  \href{https://doi.org/10.1088/1126-6708/2004/04/021}{\emph{JHEP} {\bfseries
  04} (2004) 021} [\href{https://arxiv.org/abs/hep-ph/0401119}{{\ttfamily
  hep-ph/0401119}}].

\bibitem{DeFreitas:2004kmi}
A.~De~Freitas and Z.~Bern, \emph{{Two-loop helicity amplitudes for quark-quark
  scattering in QCD and gluino-gluino scattering in supersymmetric Yang-Mills
  theory}}, \href{https://doi.org/10.1088/1126-6708/2004/09/039}{\emph{JHEP}
  {\bfseries 09} (2004) 039}
  [\href{https://arxiv.org/abs/hep-ph/0409007}{{\ttfamily hep-ph/0409007}}].

\bibitem{Ahmed:2019qtg}
T.~Ahmed, J.~Henn and B.~Mistlberger, \emph{{Four-particle scattering
  amplitudes in QCD at NNLO to higher orders in the dimensional regulator}},
  \href{https://doi.org/10.1007/JHEP12(2019)177}{\emph{JHEP} {\bfseries 12}
  (2019) 177} [\href{https://arxiv.org/abs/1910.06684}{{\ttfamily
  1910.06684}}].

\bibitem{Henn:2020lye}
J.~Henn, B.~Mistlberger, V.~A. Smirnov and P.~Wasser, \emph{{Constructing d-log
  integrands and computing master integrals for three-loop four-particle
  scattering}}, \href{https://doi.org/10.1007/JHEP04(2020)167}{\emph{JHEP}
  {\bfseries 04} (2020) 167}
  [\href{https://arxiv.org/abs/2002.09492}{{\ttfamily 2002.09492}}].

\bibitem{Henn:2016jdu}
J.~M. Henn and B.~Mistlberger, \emph{{Four-Gluon Scattering at Three Loops,
  Infrared Structure, and the Regge Limit}},
  \href{https://doi.org/10.1103/PhysRevLett.117.171601}{\emph{Phys. Rev. Lett.}
  {\bfseries 117} (2016) 171601}
  [\href{https://arxiv.org/abs/1608.00850}{{\ttfamily 1608.00850}}].

\bibitem{Henn:2019rgj}
J.~M. Henn and B.~Mistlberger, \emph{{Four-graviton scattering to three loops
  in $ \mathcal{N}=8 $ supergravity}},
  \href{https://doi.org/10.1007/JHEP05(2019)023}{\emph{JHEP} {\bfseries 05}
  (2019) 023} [\href{https://arxiv.org/abs/1902.07221}{{\ttfamily
  1902.07221}}].

\bibitem{Caola:2020dfu}
F.~Caola, A.~von Manteuffel and L.~Tancredi, \emph{{Diphoton Amplitudes in
  Three-Loop Quantum Chromodynamics}},
  \href{https://doi.org/10.1103/PhysRevLett.126.112004}{\emph{Phys. Rev. Lett.}
  {\bfseries 126} (2021) 112004}
  [\href{https://arxiv.org/abs/2011.13946}{{\ttfamily 2011.13946}}].

\bibitem{Peraro:2019cjj}
T.~Peraro and L.~Tancredi, \emph{{Physical projectors for multi-leg helicity
  amplitudes}}, \href{https://doi.org/10.1007/JHEP07(2019)114}{\emph{JHEP}
  {\bfseries 07} (2019) 114}
  [\href{https://arxiv.org/abs/1906.03298}{{\ttfamily 1906.03298}}].

\bibitem{Peraro:2020sfm}
T.~Peraro and L.~Tancredi, \emph{{Tensor decomposition for bosonic and
  fermionic scattering amplitudes}},
  \href{https://doi.org/10.1103/PhysRevD.103.054042}{\emph{Phys. Rev. D}
  {\bfseries 103} (2021) 054042}
  [\href{https://arxiv.org/abs/2012.00820}{{\ttfamily 2012.00820}}].

\bibitem{tHooft:1972tcz}
G.~'t~Hooft and M.~J.~G. Veltman, \emph{{Regularization and Renormalization of
  Gauge Fields}},
  \href{https://doi.org/10.1016/0550-3213(72)90279-9}{\emph{Nucl. Phys.}
  {\bfseries B44} (1972) 189}.

\bibitem{Heller:2020owb}
M.~Heller, A.~von Manteuffel, R.~M. Schabinger and H.~Spiesberger, \emph{{Mixed
  EW-QCD two-loop amplitudes for $q\bar{q} \to \ell^+\ell^-$ and $\gamma_5$
  scheme independence of multi-loop corrections}},
  \href{https://doi.org/10.1007/JHEP05(2021)213}{\emph{JHEP} {\bfseries 05}
  (2021) 213} [\href{https://arxiv.org/abs/2012.05918}{{\ttfamily
  2012.05918}}].

\bibitem{Chen:2019wyb}
L.~Chen, \emph{{A prescription for projectors to compute helicity amplitudes in
  D dimensions}},  \href{https://arxiv.org/abs/1904.00705}{{\ttfamily
  1904.00705}}.

\bibitem{Nogueira:1991ex}
P.~Nogueira, \emph{{Automatic Feynman graph generation}},
  \href{https://doi.org/10.1006/jcph.1993.1074}{\emph{J.Comput.Phys.}
  {\bfseries 105} (1993) 279}.

\bibitem{Vermaseren:2000nd}
J.~Vermaseren, \emph{{New features of FORM}},
  \href{https://arxiv.org/abs/math-ph/0010025}{{\ttfamily math-ph/0010025}}.

\bibitem{Studerus:2009ye}
C.~Studerus, \emph{{Reduze-Feynman Integral Reduction in C++}},
  \href{https://doi.org/10.1016/j.cpc.2010.03.012}{\emph{Comput.Phys.Commun.}
  {\bfseries 181} (2010) 1293}
  [\href{https://arxiv.org/abs/0912.2546}{{\ttfamily 0912.2546}}].

\bibitem{vonManteuffel:2012np}
A.~von Manteuffel and C.~Studerus, \emph{{Reduze 2 - Distributed Feynman
  Integral Reduction}},  \href{https://arxiv.org/abs/1201.4330}{{\ttfamily
  1201.4330}}.

\bibitem{Laporta:2001dd}
S.~Laporta, \emph{{High precision calculation of multiloop Feynman integrals by
  difference equations}},
  \href{https://doi.org/10.1016/S0217-751X(00)00215-7}{\emph{Int.J.Mod.Phys.}
  {\bfseries A15} (2000) 5087}
  [\href{https://arxiv.org/abs/hep-ph/0102033}{{\ttfamily hep-ph/0102033}}].

\bibitem{vonManteuffel:2016xki}
A.~von Manteuffel and R.~M. Schabinger, \emph{{Quark and gluon form factors to
  four-loop order in QCD: the $N_f^3$ contributions}},
  \href{https://doi.org/10.1103/PhysRevD.95.034030}{\emph{Phys. Rev.}
  {\bfseries D95} (2017) 034030}
  [\href{https://arxiv.org/abs/1611.00795}{{\ttfamily 1611.00795}}].

\bibitem{Schabinger:2011dz}
R.~M. Schabinger, \emph{{A New Algorithm For The Generation Of
  Unitarity-Compatible Integration By Parts Relations}},
  \href{https://doi.org/10.1007/JHEP01(2012)077}{\emph{JHEP} {\bfseries 01}
  (2012) 077} [\href{https://arxiv.org/abs/1111.4220}{{\ttfamily 1111.4220}}].

\bibitem{Agarwal:2020dye}
B.~Agarwal, S.~P. Jones and A.~von Manteuffel, \emph{{Two-loop helicity
  amplitudes for $gg \to ZZ$ with full top-quark mass effects}},
  \href{https://doi.org/10.1007/JHEP05(2021)256}{\emph{JHEP} {\bfseries 05}
  (2021) 256} [\href{https://arxiv.org/abs/2011.15113}{{\ttfamily
  2011.15113}}].

\bibitem{Catani:1998bh}
S.~Catani, \emph{{The Singular behavior of QCD amplitudes at two loop order}},
  \href{https://doi.org/10.1016/S0370-2693(98)00332-3}{\emph{Phys.Lett.}
  {\bfseries B427} (1998) 161}
  [\href{https://arxiv.org/abs/hep-ph/9802439}{{\ttfamily hep-ph/9802439}}].

\bibitem{Korchemsky:1987wg}
G.~P. Korchemsky and A.~V. Radyushkin, \emph{{Renormalization of the Wilson
  Loops Beyond the Leading Order}},
  \href{https://doi.org/10.1016/0550-3213(87)90277-X}{\emph{Nucl. Phys. B}
  {\bfseries 283} (1987) 342}.

\bibitem{Moch:2004pa}
S.~Moch, J.~A.~M. Vermaseren and A.~Vogt, \emph{{The Three loop splitting
  functions in QCD: The Nonsinglet case}},
  \href{https://doi.org/10.1016/j.nuclphysb.2004.03.030}{\emph{Nucl. Phys. B}
  {\bfseries 688} (2004) 101}
  [\href{https://arxiv.org/abs/hep-ph/0403192}{{\ttfamily hep-ph/0403192}}].

\bibitem{Vogt:2004mw}
A.~Vogt, S.~Moch and J.~A.~M. Vermaseren, \emph{{The Three-loop splitting
  functions in QCD: The Singlet case}},
  \href{https://doi.org/10.1016/j.nuclphysb.2004.04.024}{\emph{Nucl. Phys. B}
  {\bfseries 691} (2004) 129}
  [\href{https://arxiv.org/abs/hep-ph/0404111}{{\ttfamily hep-ph/0404111}}].

\bibitem{Grozin:2014hna}
A.~Grozin, J.~M. Henn, G.~P. Korchemsky and P.~Marquard, \emph{{Three Loop Cusp
  Anomalous Dimension in QCD}},
  \href{https://doi.org/10.1103/PhysRevLett.114.062006}{\emph{Phys. Rev. Lett.}
  {\bfseries 114} (2015) 062006}
  [\href{https://arxiv.org/abs/1409.0023}{{\ttfamily 1409.0023}}].

\bibitem{Henn:2019swt}
J.~M. Henn, G.~P. Korchemsky and B.~Mistlberger, \emph{{The full four-loop cusp
  anomalous dimension in $\mathcal{N}=4$ super Yang-Mills and QCD}},
  \href{https://doi.org/10.1007/JHEP04(2020)018}{\emph{JHEP} {\bfseries 04}
  (2020) 018} [\href{https://arxiv.org/abs/1911.10174}{{\ttfamily
  1911.10174}}].

\bibitem{Huber:2019fxe}
T.~Huber, A.~von Manteuffel, E.~Panzer, R.~M. Schabinger and G.~Yang,
  \emph{{The four-loop cusp anomalous dimension from the $N=4$ Sudakov form
  factor}}, \href{https://doi.org/10.1016/j.physletb.2020.135543}{\emph{Phys.
  Lett. B} {\bfseries 807} (2020) 135543}
  [\href{https://arxiv.org/abs/1912.13459}{{\ttfamily 1912.13459}}].

\bibitem{vonManteuffel:2020vjv}
A.~von Manteuffel, E.~Panzer and R.~M. Schabinger, \emph{{Cusp and collinear
  anomalous dimensions in four-loop QCD from form factors}},
  \href{https://doi.org/10.1103/PhysRevLett.124.162001}{\emph{Phys. Rev. Lett.}
  {\bfseries 124} (2020) 162001}
  [\href{https://arxiv.org/abs/2002.04617}{{\ttfamily 2002.04617}}].

\bibitem{Ravindran:2004mb}
V.~Ravindran, J.~Smith and W.~L. van Neerven, \emph{{Two-loop corrections to
  Higgs boson production}},
  \href{https://doi.org/10.1016/j.nuclphysb.2004.10.039}{\emph{Nucl. Phys. B}
  {\bfseries 704} (2005) 332}
  [\href{https://arxiv.org/abs/hep-ph/0408315}{{\ttfamily hep-ph/0408315}}].

\bibitem{Moch:2005id}
S.~Moch, J.~A.~M. Vermaseren and A.~Vogt, \emph{{The Quark form-factor at
  higher orders}},
  \href{https://doi.org/10.1088/1126-6708/2005/08/049}{\emph{JHEP} {\bfseries
  08} (2005) 049} [\href{https://arxiv.org/abs/hep-ph/0507039}{{\ttfamily
  hep-ph/0507039}}].

\bibitem{Moch:2005tm}
S.~Moch, J.~Vermaseren and A.~Vogt, \emph{{Three-loop results for quark and
  gluon form-factors}},
  \href{https://doi.org/10.1016/j.physletb.2005.08.067}{\emph{Phys. Lett. B}
  {\bfseries 625} (2005) 245}
  [\href{https://arxiv.org/abs/hep-ph/0508055}{{\ttfamily hep-ph/0508055}}].

\bibitem{Agarwal:2021zft}
B.~Agarwal, A.~von Manteuffel, E.~Panzer and R.~M. Schabinger, \emph{{Four-loop
  collinear anomalous dimensions in QCD and N=4 super Yang-Mills}},
  \href{https://doi.org/10.1016/j.physletb.2021.136503}{\emph{Phys. Lett. B}
  {\bfseries 820} (2021) 136503}
  [\href{https://arxiv.org/abs/2102.09725}{{\ttfamily 2102.09725}}].

\bibitem{Maitre:2005uu}
D.~Maitre, \emph{{HPL, a mathematica implementation of the harmonic
  polylogarithms}},
  \href{https://doi.org/10.1016/j.cpc.2005.10.008}{\emph{Comput. Phys. Commun.}
  {\bfseries 174} (2006) 222}
  [\href{https://arxiv.org/abs/hep-ph/0507152}{{\ttfamily hep-ph/0507152}}].

\bibitem{Cascioli:2011va}
F.~Cascioli, P.~Maierhofer and S.~Pozzorini, \emph{{Scattering Amplitudes with
  Open Loops}},
  \href{https://doi.org/10.1103/PhysRevLett.108.111601}{\emph{Phys.Rev.Lett.}
  {\bfseries 108} (2012) 111601}
  [\href{https://arxiv.org/abs/1111.5206}{{\ttfamily 1111.5206}}].

\bibitem{Buccioni:2019sur}
F.~Buccioni, J.-N. Lang, J.~M. Lindert, P.~Maierh\"ofer, S.~Pozzorini, H.~Zhang
  et~al., \emph{{OpenLoops 2}},
  \href{https://doi.org/10.1140/epjc/s10052-019-7306-2}{\emph{Eur. Phys. J. C}
  {\bfseries 79} (2019) 866}
  [\href{https://arxiv.org/abs/1907.13071}{{\ttfamily 1907.13071}}].

\bibitem{Cullen:2010jv}
G.~Cullen, M.~Koch-Janusz and T.~Reiter, \emph{{Spinney: A Form Library for
  Helicity Spinors}},
  \href{https://doi.org/10.1016/j.cpc.2011.06.007}{\emph{Comput. Phys. Commun.}
  {\bfseries 182} (2011) 2368}
  [\href{https://arxiv.org/abs/1008.0803}{{\ttfamily 1008.0803}}].

\bibitem{Bauer:2000cp}
C.~W. Bauer, A.~Frink and R.~Kreckel, \emph{{Introduction to the GiNaC
  framework for symbolic computation within the C++ programming language}},
  \href{https://doi.org/10.1006/jsco.2001.0494}{\emph{J.Symb.Comput.}
  {\bfseries 33} (2002) 1} [\href{https://arxiv.org/abs/cs/0004015}{{\ttfamily
  cs/0004015}}].

\bibitem{cln}
B.~Haible and R.~B. Kreckel, \emph{{CLN: Class Library for Numbers}}.
  \url{http://www.ginac.de/CLN}.

\bibitem{Vollinga:2004sn}
J.~Vollinga and S.~Weinzierl, \emph{{Numerical evaluation of multiple
  polylogarithms}},
  \href{https://doi.org/10.1016/j.cpc.2004.12.009}{\emph{Comput.Phys.Commun.}
  {\bfseries 167} (2005) 177}
  [\href{https://arxiv.org/abs/hep-ph/0410259}{{\ttfamily hep-ph/0410259}}].

\bibitem{Anastasiou:2000mf}
C.~Anastasiou, T.~Gehrmann, C.~Oleari, E.~Remiddi and J.~Tausk, \emph{{The
  Tensor reduction and master integrals of the two loop massless crossed box
  with lightlike legs}},
  \href{https://doi.org/10.1016/S0550-3213(00)00251-0}{\emph{Nucl. Phys. B}
  {\bfseries 580} (2000) 577}
  [\href{https://arxiv.org/abs/hep-ph/0003261}{{\ttfamily hep-ph/0003261}}].

\bibitem{Catani:1996vz}
S.~Catani and M.~Seymour, \emph{{A General algorithm for calculating jet
  cross-sections in NLO QCD}},
  \href{https://doi.org/10.1016/S0550-3213(96)00589-5}{\emph{Nucl.Phys.}
  {\bfseries B485} (1997) 291}
  [\href{https://arxiv.org/abs/hep-ph/9605323}{{\ttfamily hep-ph/9605323}}].

\end{thebibliography}\endgroup
